\documentclass[twocolumn]{aastex6}
\begin{document}


\title{WIYN OPEN CLUSTER STUDY.LXXVI. Li EVOLUTION AMONG STARS OF LOW/INTERMEDIATE MASS: THE METAL-DEFICIENT OPEN CLUSTER, NGC 2506}
\author{Barbara J. Anthony-Twarog}
\affil{Department of Physics and Astronomy, University of Kansas, Lawrence, KS 66045-7582, USA}
\email{bjat@ku.edu}
\author{Donald B. Lee-Brown}
\affil{Department of Physics and Astronomy, University of Kansas, Lawrence, KS 66045-7582, USA}
\email{donald@ku.edu}
\author{Constantine P. Deliyannis}
\affil{Department of Astronomy, Indiana University, Bloomington, IN 47405-7105 }
\email{cdeliyan@indiana.edu}
\and
\author{Bruce A. Twarog}
\affil{Department of Physics and Astronomy, University of Kansas, Lawrence, KS 66045-7582, USA}
\email{btwarog@ku.edu}


\begin{abstract}

HYDRA spectra of 287 stars in the field of NGC 2506 from the turnoff through the giant branch are analyzed. With previous data, 22 are identified as probable binaries; 90 more are classified as potential non-members. Spectroscopic analyses of $\sim$60 red giants and slowly rotating turnoff stars using line equivalent widths and a neural network approach lead to [Fe/H] = -0.27 $\pm$ 0.07 (s.d.) and [Fe/H] = -0.27 $\pm$ 0.06 (s.d.), respectively. Li abundances are derived for 145 probable single-star members, 44 being upper limits. Among turnoff stars outside the Li-dip, A(Li) = 3.04 $\pm$ 0.16 (s.d.), with no trend with color, luminosity, or rotation speed. Evolving from the turnoff across the subgiant branch, there is a well-delineated decline to A(Li) $\sim$1.25 at the giant branch base, coupled with the rotational spindown from between $\sim$20 and 70 km s$^{-1}$ to less than 20 km s$^{-1}$ for stars entering the subgiant branch and beyond. A(Li) remains effectively constant from the giant branch base to the red giant clump level. A new member above the clump redefines the path of the first-ascent red giant branch; its Li is 0.6 dex below the first-ascent red giants. With one exception, all post-He-flash stars have upper limits to A(Li), at or below the level of the brightest first-ascent red giant. The patterns are in excellent qualitative agreement with the model predictions for low/intermediate-mass stars which undergo rotation-induced mixing at the turnoff and subgiant branch, first dredge-up, and thermohaline mixing beyond the red giant bump.

\end{abstract}





\section{Introduction}

This paper presents a spectroscopic complement to the intermediate-band photometric analysis of the old open cluster, 
NGC 2506 (\citet{AT16}, hereinafter AT16). The background history of this cluster is explored in detail in AT16 and 
will not be repeated here except when required by the specifics of the analysis. Suffice it to say that precision photometry 
on the extended Str\"omgren system ($uvbyCa$H$\beta$) has established that the cluster is subject to 
low reddening ($E(B-V) = 0.058 \pm 0.001$) with negligible variation across its face (AT16). More important for our purposes, as derived from two photometric indices, the cluster is moderately metal-poor ([Fe/H] $= -0.32 \pm 0.03$) relative to the typical cluster in the solar neighborhood, a not unexpected result for a cluster positioned beyond the solar circle in the anticenter region \citep[e.g.,][]{TW97, FR02, NE16}. The key element, however, enhancing its role is its age, now well established at 1.85 $\pm$ 0.05 Gyr based upon application of the Victoria-Regina {\bf (VR)} isochrones \citep{VA06} on the Str\"omgren system. The revised cluster age places NGC 2506 in a 
category comparable to NGC 3680 at 1.7 Gyr \citep{AT09}, but slightly older than NGC 752 at 1.45 Gyr \citep{TW15} and 
IC 4651 at 1.5 Gyr \citep{AT09}; these clusters cover a range in metallicity of $\sim$0.45 dex in [Fe/H]. 

Since the overarching goal of our cluster program is to evaluate and understand the role of mixing/convection in the atmospheres and interiors of stars of low to intermediate mass both on and off the main sequence, NGC 2506 holds a pivotal position between the younger clusters like NGC 7789 (1.4 Gyr) and NGC 752 and the clearly older example of NGC 6819 (2.25 Gyr) \citep{LB15, DE18}. The slightly lower turnoff mass of NGC 2506 due to the higher age and lower [Fe/H], coupled with the richness of the cluster, allows delineation of the subgiant branch between the turnoff and the first-ascent red giant (FRG) branch below the red giant clump to a degree previously impossible, even with the combined samples of NGC 3680, NGC 752 and IC 4651 \citep{AT09}.

The outline of the paper is as follows: Sec. 2 discusses the compilation of potential spectroscopic candidates within NGC 2506, with special emphasis on the subgiant branch and stars fainter than the red giant clump, and the HYDRA observations that form
the core of this study; Sec. 3 uses radial velocities and proper motions, when available, to identify and isolate the most 
probable members, whether single stars or binaries, and to eliminate likely field stars from the analysis. Sec. 4 lays out the 
metallicity derivation of the cluster for key elements adopting a traditional technique based upon the equivalent widths of lines 
in a modest wavelength region centered on the Li 6708 \AA\ line, as well as a new approach for temperature and metallicity
estimation built around a neural network. Sec. 5 contains the derivation of the Li abundance and demonstrates 
the critical place of NGC 2506 in probing the role of mixing/convection for stars 
evolving from the main sequence to the giant branch. Sec. 6 is a summary of our conclusions.

\section{Observations and Data Reduction}
\subsection{Sample Selection}

The starting point for the selection of potential cluster members is the proper-motion study of \citet{CA81}. There is 
little doubt that the stars defined as 70\% or higher probable members based upon proper motion alone form a well-defined sample which nicely delineates the primary cluster sequences in both broad-band \citep{MC81, MA97, KI01} and intermediate-band (AT16) photometry. The weaknesses in relying solely upon this selection criterion are apparent: (a) even with a rather high proper-motion probability cutoff of 70\%, field star contamination can occur in the absence of the third vector component, the radial velocity; (b) proper-motion precision declines with increasing $V$, leading to the elimination of a higher fraction of true members and contamination by more non-members at the turnoff and fainter; and (c) the proper-motion and original broad-band 
surveys extend radially to only $\sim$5$\arcmin$ from the cluster center. Though the dominant majority of the cluster members should be contained within this zone \citep{LE13}, the desire to populate and survey the color-magnitude diagram (CMD) as completely as possible, especially along some of the relatively rapid phases of evolution beyond the the main sequence, requires a broader search to identify as many NGC 2506 inhabitants as possible. 

\subsection{Photometric Input}

To expand the sample, use was made of two CCD surveys covering comparable fields-of-view, i.e. 20$\arcmin$ on a side. 
Three broad-band CCD surveys of NGC 2506 have been undertaken to date by \citet{MA97, KI01, LE12}; the surveys cover too small an area, suffer from larger than acceptable photometric scatter, or remain unpublished, respectively.
The broad-band data for the current discussion \citep{VS10} were obtained on 6 November 2004 with the S2KB CCD at the f/7.5 focus of the WIYN 0.9m telescope at Kitt Peak National Observatory, with a scale of $0.60\arcsec$ per pixel, and seeing in the range $1.3\arcsec - 1.6\arcsec$. Cluster images were reduced using DAOPHOT II \citep{ST86}. Typically, a few hundred bright, isolated stars were used to determine a point-spread-function, which was allowed to vary spatially, and a subsequent small (typically 0.01 - 0.02 mag) aperture correction, which was also allowed to vary spatially. After rejection of outliers, approximately 50 \citet{LA92} standards for each filter provided a calibration onto the Johnson-Cousins-Landolt system, and verified the photometricity of the night, as defined by zero-point errors in the transformations below 0.01 mag. The intermediate-band photometry (AT16) came from the 
$4000 \times 4000$ CCD camera\footnote{http://www.astronomy.ohio-state.edu/Y4KCam/detector} on the 1.0m telescope operated by 
the SMARTS\footnote{http://www.astro.yale.edu/smarts} consortium at Cerro Tololo Inter-American Observatory. 
All stars with membership probabilities greater than 90\% were used to define mean relations from the main sequence 
at $V$ brighter than 15.6 to the top of the giant branch. Stars whose broad-band photometric errors placed them 
within the range of the mean relations and had proper-motion probabilities above 50\% were identified and initially 
selected. For stars outside the spatial range of the astrometric survey, location within range of the CMD relations 
provided the sole criterion for potential followup observations. 

\subsection{Spectroscopic Observations} 

Spectroscopic data were obtained using the WIYN 3.5m telescope\footnote{The WIYN Observatory was a joint facility 
of the University of Wisconsin-Madison, Indiana University, Yale University, and the National Optical Astronomy Observatory.} 
and the HYDRA multi-object spectrograph over 19 nights from 16/17 January 2015 to 24/25 February, 2017. Nine 
configurations were 
designed to position fibers on a total of 287 stars, with 38 stars observed in more than one 
fiber configuration. 
Individual exposures ranged from 10 to 90 minutes, with accumulated totals of two hours for stars 
in the brightest two configurations targeting red giants, 5 to 7.5 hours for the four configurations aimed at subgiant 
stars, and between 9 and 12.5 hours total for three constructed to sample the turnoff region and upper main sequence 
of the cluster.

 In these 58 hours of observing over a 25 month period, we obtained spectra with signal-to-noise per 
pixel $\geq 100$ for all 287 stars. 
Our spectra cover a wavelength range $\sim$400 \AA\ wide centered on 6650 \AA\ with per-pixel resolution of 0.2 \AA\ and a resolution of $R \sim$ 13000. Details on the reduction procedure can be found in \citet{LB15} and will not be 
repeated here. 

\floattable
\begin{deluxetable}{rrrcrrrrrrch}
\tablecaption{Velocity and Membership Information for Spectroscopic Sample in NGC 2506 \label{tab:table}}
\tablecolumns{12}
\tablenum{1}
%\tabletypesize\tiny
%\tablewidth{0pc}
%\setlength{\tabcolsep} {0.03in}
\tablehead{
\colhead{ID No.} & \colhead{$\alpha(2000)$} & \colhead{$\delta(2000)$} &
\colhead{Prob($\mu$)} & \colhead{$V_{rad}$} & \colhead{$\sigma(V_{rad}$)} &
\colhead{$V_{rot}sini$} & \colhead{$\sigma(V_{rot}sini$)} &
\colhead{$(B-V)$} & \colhead{$V$} & \colhead{Membership} & \nocolhead{PhilNo}
}
\startdata 
7001 & 119.838542 & -10.705722 & .... & 50.1 & 0.8 & 18.0 & 0.4 & 0.886 & 14.990 & NM & 3693 \\
7002 & 119.839917 & -10.872389 & .... & 44.5 & 0.8 & 19.9 & 0.5 & 0.988 & 12.494 & NM & 636 \\
7003 & 119.842792 & -10.805722 & .... & -23.5 & 1.5 & 28.6 & 1.2 & 0.464 & 14.859 & NM & 1648 \\
7004 & 119.843833 & -10.855722 & .... & 61.0 & 0.7 & 17.3 & 0.4 & 0.856 & 14.578 & NM & 349 \\
7005 & 119.845708 & -10.889056 & .... & 60.6 & 0.1 & 16.2 & 1.7 & 1.030 & 13.466 & NM & 213 \\
7006 & 119.850667 & -10.822389 & .... & 27.0 & 2.6 & 33.2 & 2.0 & 0.322 & 14.689 & NM & 1228 \\
7007 & 119.852095 & -10.891248 & .... & -13.5 & 1.6 & 51.1 & 1.3 & 0.890 & 11.600 & NM & 458 \\
7008 & 119.852125 & -10.839056 & .... & 83.6 & 0.7 & 9.1 & 0.2 & 0.897 & 14.703 & M & 1060 \\
7009 & 119.855000 & -10.789056 & .... & 74.6 & 0.7 & 20.3 & 0.5 & 0.642 & 14.495 & NM & 1897 \\
7010 & 119.857170 & -10.924294 & .... & 103.7 & 2.3 & 29.2 & 2.7 & 0.394 & 15.210 & NM & 96 \\
  &   &   &  &  & & & & & & & \\
7011 & 119.861862 & -10.760853 & .... & 101.5 & 0.7 & 11.0 & 0.3 & 1.071 & 13.935 & NM & 2656 \\
7012 & 119.862495 & -10.905938 & .... & 69.4 & 0.7 & 11.3 & 0.3 & 1.061 & 13.904 & NM & 290 \\
7013 & 119.862595 & -10.620172 & .... & 39.3 & 1.1 & 18.2 & 0.6 & 1.304 & 12.546 & NM & 4865 \\
7014 & 119.864540 & -10.904671 & .... & 23.1 & 0.6 & 9.7 & 0.2 & 0.600 & 14.529 & NM & 152 \\
7015 & 119.866119 & -10.894622 & .... & 83.8 & 2.7 & 85.5 & 6.6 & 0.399 & 15.174 & M & 400 \\
7016 & 119.871458 & -10.939056 & .... & -45.9 & 1.0 & 14.2 & 0.5 & 0.575 & 14.335 & NM & 7 \\
7017 & 119.872849 & -10.654552 & .... & 59.8 & 2.4 & 60.7 & 3.3 & 0.439 & 14.419 & NM & 4456 \\
7018 & 119.879631 & -10.691129 & .... & 27.0 & 0.5 & 13.7 & 0.2 & 0.873 & 12.398 & NM & 3983 \\
7019 & 119.880058 & -10.631248 & .... & 82.9 & 0.6 & 9.1 & 0.2 & 0.910 & 14.548 & M & 4732 \\
7020 & 119.880890 & -10.916945 & .... & 20.1 & 0.6 & 14.8 & 0.3 & 0.782 & 14.730 & NM & 172 \\
\enddata
\tablecomments{A short sample of lines is presented here to indicate the format and content of the full table which will be available online.}
\end{deluxetable}

Multiple exposures of any particular fiber configuration were combined if the observations were obtained 
within the same run of a few adjacent nights.
 A few of our configurations were observed over a period of 
a year or more, with some fiber losses in the interim. In these cases, the combination of individual spectra for 
stars obtained through different fibers or in different years was accomplished by undoing the individual 
throughput corrections before combining the spectra. This was also the procedure for stars observed in 
different runs as part of different fiber configurations. Combination of spectra from
 widely separated 
epochs was only carried out if there appeared to be no significant velocity shift.

As always, inclusion of specific stars within the spectroscopic survey was ultimately constrained by the
need to optimize HYDRA configurations over multiple observing runs. As preliminary reductions revealed 
that some stars were clear radial-velocity non-members, these were dropped from the program and others, 
usually potential subgiants and giants, were added. A small set of definite non-members was included in 
the data base to use as a control for comparison with the hopefully homogeneous group of metal-deficient cluster members.

Before discussing the results for individual stars, it should be noted that, whenever available, a star will be referred to
using its WEBDA identification. However, since the sample here covers a much broader area than that of past published
surveys, regrettably another identification number must be added for those stars outside the area listed within WEBDA. 
Column 1 of Table 1 lists WEBDA identification numbers for stars; numbers greater than 7000 refer to stars added by
this survey. Table 1 lists stars in sequence based upon their right ascension; coordinates given in Table 1 are 
identical to those found in AT16 for all but 10 stars not included in the photometric study. The typeset 
version of Table 1 includes enough lines to 
show the form and content of the larger table available online.

A point of confusion regarding the WEBDA identifications does require correction. In the original photometric 
and astrometric surveys \citep{MC81, CA81}, the stars were numbered using a quadrant number, ring number, and sequential
count within that zone. For quadrant 2, ring 3, the number of stars extended to 106, so the last star measured was 23106.
WEBDA incorrectly rewrote the numbers above 100 as ring 4, i.e. 23106 became 2406. This leads to confusion because in quadrants
2 and 4, there are photoelectric standards outside the three rings of the photographic survey numbered 2401, 2402, 4401, and 4402.
WEBDA 2401 and 2402 are actually 23101 and 23102. In our Table and analysis, 2401 and 2402 refer to the original
designations as marked on the cluster chart (Fig. 1) of \citet{MC81}.

\subsection{Radial Velocities and Rotation}

Individual heliocentric stellar radial velocities, $V_{rad}$, were derived from each summed composite spectrum 
utilizing the Fourier-transform, cross-correlation facility {\it fxcor} in IRAF \footnote{IRAF is distributed by the National Optical Astronomy Observatory, which is operated by the Association of 
Universities for Research in Astronomy, Inc., under cooperative agreement with the National Science Foundation.}. 
In this utility, program stars are compared to stellar templates of similar 
effective temperature ($T_{\mathrm{eff}}$) over the full wavelength range of our spectra excluding the immediate 
vicinity of the H$\alpha$ line. Output of the {\it fxcor} utility characterizes the cross correlation function, 
from which estimates of each star's radial velocity are easily inferred. 
Rotational velocities can also be estimated from the cross correlation function full-width (CCF FWHM) using a procedure developed by \citet{ST03}.This procedure exploits the relationship between the CCF FWHM, line widths and $V sin i$, using a set of numerically ``spun up" standard spectra with comparable spectral types to constrain the relationship.   
For a significant fraction of our sample, higher rotational 
velocities conspired with weaker lines to produce spurious values 
of the radial velocity when the region near H$\alpha$ was excluded; 
quoted values in our table include stars for 
which the cross correlation included H$\alpha$.
 For simplicity, $V_{rot}$ as used here implicitly includes the unknown $sin$ $i$ term. 

There are two means to test the precision of the radial velocities, the size of the scatter among multiple observations of
the same star and the scatter in residuals relative to an independent source of measurements. Over the multiple HYDRA runs,
36 stars were observed twice and 2 stars were observed 3 times. The dispersions in radial velocity for the latter pair 
are 0.70 and 0.55 km s$^{-1}$. Among the stars with two observations, star 7112 showed a difference of 28 km s$^{-1}$ between the two epochs and is clearly a double-lined spectroscopic binary. It sits among the giants at the level of the clump and undergoes eclipses \citep{AR07}. With both sets of lines visible, it is likely that the system is composed of two
first-ascent red giants. 

For the 35 remaining paired sets of observations, the 
average absolute difference in the radial velocities is 0.84 $\pm$ 0.96 km s$^{-1}$. However, the two pairs with the largest 
discrepancies, 3.22 and 4.62 km s$^{-1}$, are stars 1359, categorized by \citet{ME07} as a probable binary, and 4376, tagged 
as a probable binary in the current investigation due to the range in published velocities for this star. If these two 
are removed from the sample, the mean offset between paired observations becomes 0.65 $\pm$ 0.57 km s$^{-1}$. 

We next compare our data with that of \citet{ME07, ME08}, which supercedes the smaller and/or less precise samples 
of \citet{FJ93, MI95, CA04, RE12}, and with the more recent work of \citet{CA14, CA16}. Of the 30 members and 4 non-members 
observed by \citet{ME07}, all but two of the non-members and two of the members were included in our observations. 
For the remaining 30 stars, the mean residual in radial velocity, in the sense (MM - Table 1), is +0.58 $\pm$ 1.80 km s$^{-1}$.
If probable binaries 1359, 2251, and 4376 are removed, the mean residual drops to +0.31 $\pm$ 0.96 km s$^{-1}$. 
It is encouraging to note that the dispersion is very comparable to that derived from a similar analysis of the giants 
and main sequence stars in NGC 6819 \citep{LB15}. Our cluster mean velocity for the 25 members is 83.2 $\pm$ 1.2 km s$^{-1}$.

\citet{CA14} has published radial velocities for 27 red giant members of NGC 2506; all but 2122 are included in the current study.
From the residuals between the two samples, five stars stand out as anomalous, 2109 and 2276, as well as the already 
identified 1359, 2251, and 4376. If these stars are eliminated, the 21 remaining stars generate mean residuals of +0.63 $\pm$ 1.04. \citet{CA14} finds a mean radial velocity for these 21 giants of +83.9 $\pm$ 1.1 km s$^{-1}$.

Finally, \citet{CA16} present observations of five red giants in NGC 2506, all of which were included in the three previous studies. While velocities of four of the stars are consistent with previous observations within the uncertainties, star 3265 exhibits a range from 80.3 to 85.3 km s$^{-1}$. While not conclusive, the spread could be indicative of a binary classification, as noted by \citet{CA16}, who also emphasize the peculiar location of this star in the CMD and the distinctly anomalous [Fe/H] derived relative to the other four giants.

\section{Cluster Membership}

\subsection{Probable Binaries}
To minimize the potential distortions in any trends between main sequence and red giant stars, in addition to removing non-members,
it's valuable to identify likely binaries which may have either anomalous colors and spectra, as well as modified evolution due
to binary interaction. Among the giants, it has already been noted that stars 1359, 2109, 2251, 2276, 3265, and 4376 are probable
binaries. \citet{AR07} have used intermediate-band photometry to identify a large sample of variables in the field of NGC 2506,
including 3 eclipsing binaries brighter than $V$ = 15.6. This sample includes the already noted giant, 7112, and two stars near the turnoff, 1136 and 1212, both of which appear to be radial-velocity cluster members based upon preliminary analysis of unpublished high resolution spectroscopy of these stars \citep{AR07}. Star 3255 (V10) is a red giant blueward of the clump with two almost identical radial-velocity measures, both consistent with cluster membership. \citet{AR07} find variability at the mmag level with a periodicity of 10 d$^{-1}$, though it does lie near the saturation limit of their photometric survey. As a member, it is clearly not a $\delta$ Scuti star.
 
A straightforward method for tagging potential binaries from single spectra is to look for a double set of lines within 
the spectrum, as revealed via {\it fxcor} or, at minimum, a consistent asymmetry in the line profile for a spectrum caused by the overlap of two lines. From visual and {\it fxcor} inspection of all the spectra, stars 1106, 2144, 2376, 3378, 4262, 5086, 7025, 7033, 7050, 7057, 7052, 7101 and both spectra for 7112 were noted as possible spectroscopic binaries. 

It should be mentioned that the apparently low fraction of identified binaries ($< 10$\%) is not indicative of the probable cluster fraction and is primarily tied to the combination of selection effects in the initial selection of probable cluster members at the turnoff, the limited number of radial-velocity measures per star, and the limits on the radial-velocity measurements for stars with higher than average rotation.

\subsection{Radial-Velocity Membership}
With radial velocities for 287 stars but proper-motion membership for only 129, final membership for the majority of the sample is strongly dependent upon the single velocity component, weighted by location within the CMD. While this approach 
worked extremely well for NGC 6819 \citep{LB15}, the younger age by 0.4 Gyr places the turnoff of NGC 2506 in a regime 
where rotational speeds can remain high (greater than 25 km s$^{-1}$), significantly impacting the precision of the radial 
velocity. It should be noted that for the giants, the measured rotational velocity is dominated by the resolution 
of the spectra and therefore should be regarded as an upper limit to the true value. By contrast, \citet{CA14} derived precision rotational velocities from $R$ $\sim$ 44000 spectra from the MIKE 
spectrograph. From 21 single-star members common to our sample, \citet{CA14} finds an average rotational speed of 
3.2 $\pm$ 1.2 km s$^{-1}$; our lower $R$ data generate an average of 13.8 $\pm$ 3.0 km s$^{-1}$. 

To determine membership by radial velocity alone, any star with a radial velocity more than three $\sigma$ from the cluster 
mean as defined by our single cluster giants (83.3 km s$^{-1}$) will be classified as a probable non-member. To set the individual 
$\sigma$ for each star, two factors were taken into account. First, as expected, the calculated uncertainty in the radial velocity
is well correlated in approximately linear fashion with the rotational speed. We averaged the calculated error in radial 
velocity as defined by $fxcor$ and the error as predicted from $V_{rot}$, i.e. $\sigma_{Vrad}$ = 0.044*$V_{rot}$ + 0.246. 
Second, in the limit of perfect velocity measurements, this estimate implies no expected dispersion
among the stars, ignoring the intrinsic velocity dispersion among the sample caused by motion about the center of mass 
of the cluster. Thus, the previously quoted dispersions among the purportedly single-star radial velocities for 
the red giants in our sample (1.2 km s$^{-1}$) and that of \citet{CA14} (1.1 km s$^{-1}$) are combinations of both measurement errors and the intrinsic spread among the giants. The slight improvement in the dispersion for the data with higher resolution isn't unexpected so, as an approximate estimate for the intrinsic radial velocity spread, we 
adopt 1.0 km s$^{-1}$. This is effectively the same value we would get if we adopted 1.2 km s$^{-1}$ for the total dispersion, but removed 0.6 - 0.7 km s$^{-1}$ as the instrumental scatter derived from multiple observations of the same stars over different
runs. The final adopted $\sigma$ in the individual $V_{rad}$ becomes the addition in quadrature of 
the fixed intrinsic cluster spread with the individual measurement error as derived above.

Application of the radial velocity criterion to 265 probable non-binary stars leads to initial membership for 187 stars, though we emphasize that the claim to single-star status is based upon a lack of direct evidence for binarity, which is difficult to come by for stars with only one spectroscopic observation.
Of these 187 stars, 98 have proper-motion probabilities, 86 of which are above 50\%. As the second cut, we
eliminate the 12 stars with probability ranging from 0\% to 42\%, leading to a final sample of 175 probable, single-star
members. It is encouraging to note that of the 78 stars classified as radial-velocity non-members, 17 have proper-motion
probabilities of which 13 lie below 50\%. The distribution of single-star members (solid black line) and 
non-members (dashed blue curve) based solely on radial velocity is shown in Fig. 1.

\begin{figure}
\figurenum{1}
\plotone{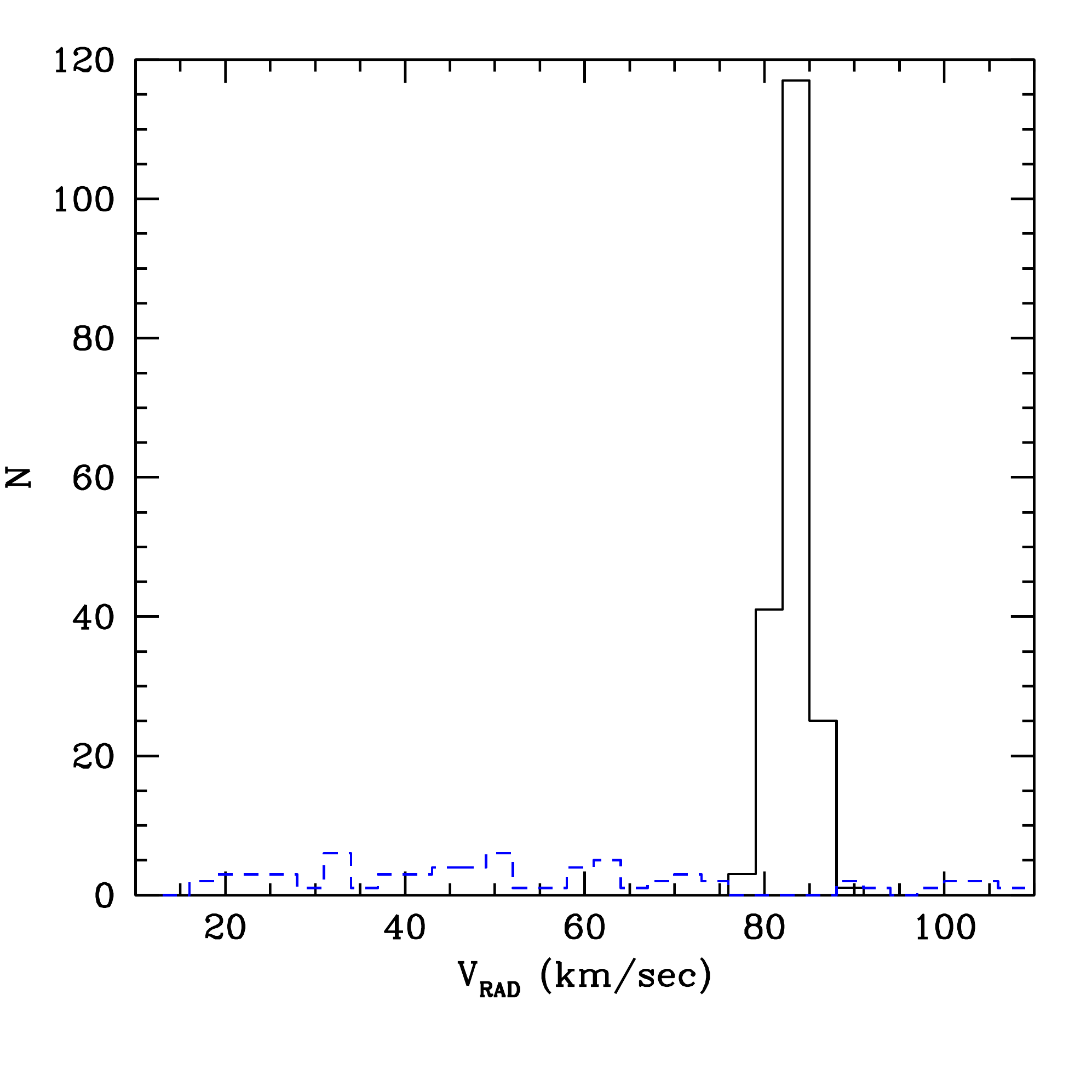}
\caption{Distribution of radial velocities for single stars classified as probable radial-velocity members (solid black line) and potential non-members from radial-velocity alone (blue dashed line).}
\end{figure}

\subsection{The Color-Magnitude Diagram}

To optimize the precision of the photometric input for the atmospheric parameters needed for the spectroscopic analysis, 
particularly $T_{\mathrm{eff}}$ from colors and $\log g$ from location within the CMD, use will be made of three 
photometric indices, $b-y$, $B-V$, and $hk$. For the color temperature, the $B-V$ values of \citet[][hereafter VS]{VS10} defined 
the starting point. As a check on the photometric zero-point, comparison with the photoelectric observations of \citet{MC81} 
for 29 stars, 23 brighter than $V$ = 16.0, led to a mean offset
of -0.010 $\pm$ 0.025, in the sense (MC - VS), which was applied to the CCD $B-V$ system. 

Photometric indices from the intermediate-band (AT16) and broad-band (VS) surveys are available for 
over 2000 stars. A cubic relation between $(b-y)$ and $(B-V)$ was determined using a set of 730 stars with the 
lowest color errors; the relationship has a standard deviation in $(B-V)$ color of 0.022. A similar relationship 
was determined between $hk$ indices and $B-V$ using solely cool member giants (AT16) to provide an additional 
temperature measure.

The spectroscopic sample includes ten stars without Str\"omgren photometry due to placement slightly outside the 
field of the published Str\"omgren survey. For nine of these stars, the VS photometry was used exclusively. 
Conversely, the broad-band photometry suffered a few gaps, particularly at the very bright end. 
For four of the five stars lacking photometry from the VS survey, we relied exclusively on 
the Str\"omgren photometry. These few very bright stars targeted primarily the red giants, although one of 
the five bright stars, 1375, is clearly a hot star. For the one star lacking photometric indices from both 
surveys, 7007 (Tyc 5416-2526-1), Tycho values were used for $V$ and $B-V$.

From 273 nonvariable stars in the spectroscopic survey in common with VS, the mean difference in $V$ in the 
sense (AT16 - VS) is -0.009 $\pm$ 0.021. Since the AT16 $V$ system is in excellent agreement with that of \citet{MC81}, where
the mean difference in $V$ from 33 stars brighter than $V$ = 17.0 is -0.001 $\pm$ 0.034 in the sense (MC - AT16),
a simple average of AT16 and VS was adopted for $V$.
For 9 of the 10 stars not surveyed by AT16, the $V$ magnitudes of VS were adopted without modification.

Fig. 2a shows the CMD for all stars observed spectroscopically; Fig. 2b is composed of NGC 2506 radial-velocity and, if 
available, proper-motion members for which no evidence of binarity currently exists. Table 1 summarizes the basic information
about all stars observed spectroscopically. Stars listed as M are probable radial-velocity members, NM are nonmembers, MB are probable binary 
members, MN are radial-velocity members  with proper-motion membership below 50\%, B are probable binaries for 
which the radial velocity deviates significantly from the mean, and BNM and BM are binaries with deviant radial velocities
but proper-motion probabilities below and above 50\%, respectively. 

What is apparent from a comparison of the two CMDs is that the greatest decline in stars due to the elimination of radial-velocity
non-members is concentrated in two regions of the CMD. The first is near the top of the vertical turnoff, between $V$ = 14.2 and 14.8. A large sample from outside the original proper-motion survey region was selected from this region of the CMD in the hope of 
mapping the turnoff hook and the inital phase of evolution to the subgiant branch. The second concentration is among the 
red giants in the vertical band between $B-V$ = 0.8 and 1.2, where stars were initially retained in the hope of adding to 
the cluster subgiant and red giant branches. While some of these stars, particularly brighter than the clump, proved to be 
non-members, 29 probable subgiant and giants members were added to the final sample. Again, both color regions will be 
contaminated by field stars at any age younger than $\sim$3 Gyr for the turnoff and at almost any age for disk red giants 
and clump stars. Since we only have one velocity component for these stars, even with relatively tight radial-velocity limits,
it is probable that a handful of field non-members are still contained within the member sample.

\begin{figure}
\figurenum{2}
\plotone{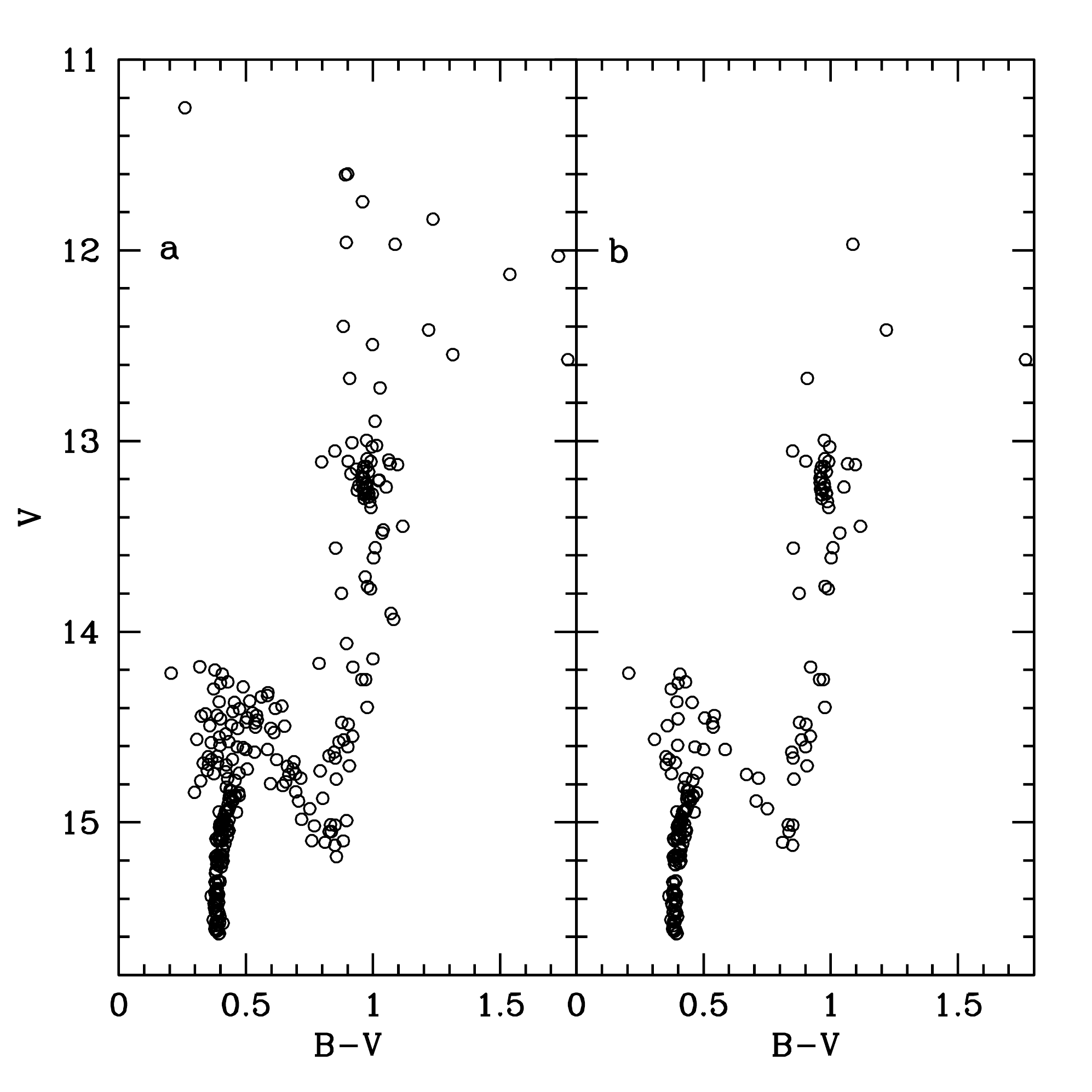}
\caption{(a) CMD for all stars observed spectroscopically. (b) Probable single-star members of the spectroscopic sample.}
\end{figure}

\section{Metallicity}
Determination of stellar elemental abundances from high dispersion spectroscopy depends upon a variety of factors, from the signal-to-noise
ratio of the spectroscopic region under analysis to the fundamental stellar parameters defining the strength and appearance of the line profiles:  $V_{rot}$, 
$T_{\mathrm{eff}}$, $\log g$, and microturbulent velocity, $v_{t}$. With high resolution over an extended wavelength region and a large number of lines for both neutral and ionized elements, one can constrain the potential range of parameters by requiring a lack of trends between the derived
abundances and the fundamental parameters, as well as the excitation potential. For a smaller spectral wavelength range with fewer lines, one is
forced to fall back on alternative means of deriving the key parameters, independent of the spectrum, e.g. using a color-based $T_{\mathrm{eff}}$ and $\log g$ defined by position in the CMD for cluster members.

Since the first cluster analysis in this series for NGC 3680 \citep{AT09}, where a modest number of mostly dwarf stars and a few giants were
being considered, our approach has evolved to allow discussion of samples expanded by almost an order of magnitude compared to earlier work
by us and others. A key change, as exemplified by NGC 6819 \citep{LB15}, has been the adoption of an automated line-measurement program, 
ROBOSPECT \citep{WH13}, to replace exclusive dependence on manual measurement of line equivalent widths as input in traditional model atmosphere analysis 
via MOOG \citep{SN73}. Because ROBOSPECT consistently performed as well as manual measurements over the color range from dwarfs to giants, the primary sources of uncertainty among the final abundances within a cluster ultimately depended upon the primary parameters of $T_{\mathrm{eff}}$ and $v_{t}$. In theory, one could adjust these two scales for the stellar sample until the abundances showed no trend with either variable, but this would minimize the slope without supplying an independent check on the zero-point of the abundances. In particular, 
the color-$T_{\mathrm{eff}}$ relations adopted to date for the dwarfs and giants come from two independent relations, each of which has its own slope and zero-point, leading to the possibility of a mismatch between these two subsamples within the same cluster. While one could check the derived parameters against those already published for cluster stars observed at high dispersion by others, in many cases the samples only 
discuss either dwarfs or giants, but not both, or little if any high dispersion work is available for the cluster of interest.

As an alternative to our photometric $T_{\mathrm{eff}}$ values and EW-based spectroscopic [Fe/H] estimates, we have attempted to derive $T_{\mathrm{eff}}$ and [Fe/H] for each star in our sample using ANNA (Lee-Brown 2018, in prep), a new, flexible, Python-based code for automated stellar parameterization. ANNA utilizes a feed-forward, convolutional neural network \citep{AR02}, a machine-learning technique, to infer stellar parameters of interest from input spectra. Multiple tests show that ANNA is capable of producing accurate metallicity estimates with precision competitive with our EW-based analysis. Additionally, ANNA is capable of accurately inferring $T_{\mathrm{eff}}$ from our spectra alone, providing an alternate temperature determination for each star. A deeper discussion of ANNA's design and capabilities will be given in Lee-Brown (2018), but we briefly summarize its operation here. ANNA is freely available for download; the version of ANNA used in this investigation can be found at Zenodo, while the current version of the code can be found at GitHub. 

\subsection{ANNA: A Neural Network for Temperatures and Abundances}

ANNA builds on previous studies examining the suitability of neural networks as tools to parameterize stellar spectra \citep[e.g.,][]{BA97, AL00, SN01, MA10, DA16, LI17}. 
The neural network used in ANNA consists of {\it layers} of sequential mathematical operations. The first, or input, layer is a stellar spectrum consisting of an ordered sequence of pixel values. These values are then summed according to many different sets of weight vectors, such that a set of weighted sums of the inputs is calculated. This collection of weighted sums forms a {\it hidden} layer. Each of the weighted sums in this hidden layer then serves as input to a non-linear function. In ANNA, this function is the rectified linear unit (ReLU), defined as $f(x) = max(x, 0)$, where $x$ denotes the weighted sum serving as input to the function. The outputs from this function are then used to construct another set of weighted summations, organized into another hidden layer, and the process repeats several times. The final result of this sequence of operations is a set of outputs $\{ y_i \}$ which, in the case of ANNA, consists of the stellar parameters of interest that have been inferred from the input spectrum.

ANNA contains five layers: an input and an output layer connected via three sequential hidden layers. The first hidden layer is a convolutional layer which contains a collection of filters that are each convolved with the inputs. The filters consist of collections of weights used to construct a weighted sum that only depends on a localized subset of the layer inputs. The filters are windowed across the inputs to generate many such localized summations. This localization reduces the number of free parameters (weight values) in the first hidden layer, helping to reduce overfitting during training while improving computational performance \citep[see][]{AR02}. The other two hidden layers are fully-connected; each summation in these layers is connected via weights to all the ReLU operations in the previous layer.

ANNA's ability to translate between input spectra and accurate stellar parameters requires correct selection of each weight value used during the summation steps. This selection is done automatically through supervised learning. Initially, small weight values are sampled randomly from a normal distribution. The network is then provided an input consisting of a spectrum whose output parameters are known. ANNA will produce outputs based on the input spectrum; the accuracy of these reported outputs can be quantitatively assessed via a cost function, $J(y_{\mathrm{true}}, y_{\mathrm{infer}})$, that measures the deviations of the reported outputs from the known stellar parameters. In ANNA, a quadratic cost function is used, $J = \Sigma (y_{\mathrm{true}} - y_{\mathrm{infer}})^2$. Thus, for large deviations of the network outputs from the known values, the value of $J$ is correspondingly large. Minimization of $J$, therefore, implies that the network's weights are such that $y_{\mathrm{infer}}$ closely matches $y_{\mathrm{true}}$. This minimization is accomplished by computing the gradient of $J$ with respect to each network weight, followed by an update of the weights in the direction of lower $J$. Through successive iterations of this updating process, the weights eventually converge to optimal values. The network will then be capable of inferring stellar parameters from spectra whose parameterizations are unknown. 

\begin{deluxetable}{lr}
\tablenum{2}
\tablecaption{ANNA Training Set Parameters}
\label{Tab:annaranges}
\tablehead{\colhead{Parameter} & \colhead{Range}}
\startdata
$T_{\mathrm{eff}}$ (K) & 3500 to 7500 \\
{[}Fe/H] (dex) & -1.0 to 0.0 \\
$log(g)$ (dex) & 2.0 to 4.5 \\
$v_t$ (km s$^{-1}$) & 1.0 to 2.5 \\
$V_\mathrm{rot}$ (km s$^{-1}$) & 0 to 70 \\
$RV$ (km s$^{-1}$) & -10 to 95 \\
S/N & 100 to 250 \\
\enddata
\end{deluxetable}

We trained ANNA using continuum-normalized spectra generated with the code SPECTRUM \citep{GR94}, Kurucz atmospheric models \citep{KU92}, and a custom linelist generated using VALD \citep{KU99} that was adjusted to reproduce the solar spectrum. We generated 15,000 high-resolution ($R\sim 670,000$) training spectra with parameters randomly selected between the ranges given in Table 2. The high resolution spectra were then post-processed to better mimic our sample of HYDRA spectra. This post-processing included random radial-velocity shifts, rotational broadening, and smoothing using a Gaussian line-spread function to a resolution of $R \sim 9000$. This resolution is lower than the actual resolution of our spectra ($R \sim 13000$), and was selected during testing of ANNA as it represents the resolution corresponding to the minimum average RMS deviation between our synthetic training spectra and a sample of real Hydra spectra. This adoption of a lower-than-actual resolution for our training spectra likely indicates that our synthetic models imperfectly reproduce the line-spread function of the spectrograph and/or do not completely model the subtle broadening effects due to changes in surface gravity or microturbulent velocity. However, during testing we ultimately determined that ANNA's temperature and metallicity determinations were relatively insensitive to the adopted training resolution; our ANNA results are materially unchanged if we were to instead adopt $R \sim$ 13000 during training. This is likely due to the fact that temperature and metallicity are more sensitive to relative line strengths, rather than the particular shapes of the line profiles.

The pixel scale was set to 0.2 $\mathrm{\AA}$  px$^{-1}$. We also limited the wavelength coverage of our training spectra to 6625 \AA\  - 6825 \AA. This was done to avoid having to model calibration artifacts in our real spectra in the region of H$\alpha$. After post-processing, our training sample consisted of 225,000 spectra. 

During training, subsamples of 100 randomly selected example spectra were used during each weight update iteration. When spectra were selected for training, a wavelength-dependent amount of noise was added according to a randomly selected S/N value and a relative S/N template derived from HYDRA observations of the sun. Additionally, small, random continuum offsets were added to simulate continuum placement errors in our real spectra. Training was carried out until the network cost function failed to improve within 20,000 weight update iterations, for a total of approximately 200,000 iterations. 

After training, we verified the capabilities of the trained network using a sample of real HYDRA spectra. This sample included spectra of the sun, as well as members of the open clusters NGC 6819 \citep{LB15} and the Hyades \citep{CU17}. These test spectra were first linearly interpolated onto the wavelength grid used during the training process. Of the potential stellar parameters, the trained network most reliably determined the correct $T_{\mathrm{eff}}$ and [Fe/H]; this is unsurprising as these two parameters contribute most strongly to the observable features in our selected spectral range. Hotter stars in our test sample ($T_{\mathrm{eff}}\sim$ 6500 K) could not be accurately parameterized by ANNA, likely due to a lack of strong spectral features over the wavelength range of interest. For cooler stars, we verified that ANNA returned reliable parameter determinations down to $T_{\mathrm{eff}}\sim$ 4000 K  using NGC 6819 spectra. With the exception of one star (4402), which we omit from our ANNA analysis, the stars in our NGC 2506 sample have surface temperatures well above this tested value. 

ANNA's  $T_{\mathrm{eff}}$ and [Fe/H] results derived from the cooler ($T_{\mathrm{eff}}\leq$ 6500 K) test spectra compare favorably with parameters known \textit{a priori}. For a sample of 89 solar spectra of typical S/N = 200 obtained from the daytime sky with HYDRA for calibration purposes during our observing runs for NGC 2506, ANNA returns 
an average $T_{\mathrm{eff}} = 5780 \pm 77$ K (s.d.) and average [Fe/H] = 0.04 $\pm 0.03$ dex (s.d.). From the rich sample of 37 giants and 184 dwarfs in NGC 6819, ANNA derives [Fe/H] = -0.01 $\pm$ 0.09 (s.d.) and [Fe/H] = -0.06 $\pm$ 0.08 (s.d.), respectively, in excellent agreement with the EW analysis \citep{LB15} which generates [Fe/H] = -0.03 
$\pm$ 0.09 (s.d.). Finally, despite the Hyades' super-solar metallicity, [Fe/H] $\sim 0.15$ \citep[see discussion in][]{CU17}, which places it outside the range of metallicities used during training, ANNA correctly infers [Fe/H] = 0.17 $\pm  0.05$ (s.d.). These results collectively indicate ANNA's reliability over a range in stellar $T_{\mathrm{eff}}$ and [Fe/H] typical of disk clusters.

\subsection{ANNA Parameters: NGC 2506}
The initial database of spectra processed through ANNA consisted of the 175 dwarfs and giants classed as single-star members, as detailed in Table 1. It quickly became apparent that the stars at the turnoff, even with $T_{\mathrm{eff}}$ below 6500 K, generated results which were inconsistent and/or subject to larger than desirable  errors. Since the same conclusion was reached after using ROBOSPECT, independent of ANNA, and by tests using manual equivalent width measurements for a representative subset of turnoff stars, the source of the problem lies with the spectra rather than the technique adopted. 
In contrast with our work on NGC 6819, the fundamental weaknesses of the dwarf analyses in NGC 2506 are due to the lower metallicity of the cluster, the hotter $T_{\mathrm{eff}}$ due to the younger age combined with a lower metallicity, and the significantly wider range of $V_{rot}$. In fact, the large majority of stars at the turnoff of NGC 6819 have $V_{rot}$ below 25 km/sec \citep[][in prep.]{DE18}, in contrast with NGC 2506 where this value defines an approximate lower bound for the majority of stars at the turnoff.

We therefore used ANNA to derive $T_{\mathrm{eff}}$ and [Fe/H] only for stars with $(B-V)_0$ between 0.65 and 1.15, excluding only one cool giant
with $(B-V)_0$ greater than 1.7. 
The results for 62 red giant members are presented in Table 3 under the columns $T_{NN}$ and [Fe/H]$_{NN}$. From these 62 stars, we first construct a mean [Fe/H] value $= -0.25 \pm 0.13$ (s.d.).  The majority of the scatter is caused by three stars which have abundances ranging from solar to twice solar. These metallicities are consistent with those derived from ROBOSPECT measures and traditional equivalent width analyses. The implication is that these are, in fact, likely field stars with distinctly higher [Fe/H] than the cluster. If these 3 stars are dropped, the revised cluster metallicity average and standard deviation from ANNA analysis becomes [Fe/H] $= -0.27 \pm 0.06$ (s.d.). The median of individual stellar [Fe/H] values is robustly estimated as $-0.26\pm 0.03$ with an error estimate based on a median of star-by-star absolute deviations from the cluster [Fe/H] value.

\subsection{ROBOSPECT}
In keeping with our approach to spectroscopic abundance determination for previous clusters in this program, our default scheme for determining model atmosphere input temperatures was based upon photometric color, specifically 
$B-V$, defined in the current investigation  as the average of the observed $B-V$, $b-y$ converted to $B-V$, and, 
when available for cluster members, $hk$ converted to $B-V$ for red giant members of the cluster. As noted earlier,
the $T_{\mathrm{eff}}$ for each star had been based on two primary color-temperature calibrations. 

For dwarfs, the adopted calibration is that of \citet{DE02}, consistent with previous and ongoing spectroscopic studies 
by this group and compared in detail with more recent $T_{\mathrm{eff}}$ calibrations in \citet{CU17}, namely: 

\begin{eqnarray}
\nonumber
T_{\mathrm{eff}}(K) = 8575 - 5222.7 (B-V)_0 \\ 
+ 1380.92 (B-V)_0^2 + 701.7 (B-V)_0[{\rm [Fe/H]} - 0.15] 
\end{eqnarray}

\noindent
In previous investigations, the giant star color-temperature calibration of \citet{RA05} was used for stars with $B-V > 0.5$.
With the availability of the revised $T_{\mathrm{eff}}$ estimates produced by ANNA for subgiants and giants, we 
have replaced the cool-star calibration with the newer values. To mesh the scales smoothly, the reddening-corrected $(B-V)_0$ 
values for dwarfs have been run through the dwarf $T_{\mathrm{eff}}$ calibration for all stars bluer than $(B-V)_0$ = 0.34, 
adopting [Fe/H] = -0.30. For stars redder than $(B-V)_0$ = 0.7 to $(B-V)_0$ = 1.2, we have adopted the ANNA $T_{\mathrm{eff}}$ values,
excluding the three stars which appear to be metal-rich from both ANNA and ROBOSPECT analyses. 
These data were then fit with a cubic relation:

\begin{eqnarray}
\nonumber
T_{\mathrm{eff}}(K) = 8751.5 - 6955.7 (B-V)_0 \\ 
+ 4519.1 (B-V)_0^2 - 1461.1 (B-V)_0^3 
\end{eqnarray}

\noindent
 Since the relation is tied to a specific [Fe/H], it is appropriate for NGC 2506 alone.

From the 107 single-star dwarfs, the mean $T_{\mathrm{eff}}$ offset, in the sense (OLD - NEW), is -0.2 $\pm$ 6.5 K.
For 53 giants, the analogous comparison between ANNA $T_{\mathrm{eff}}$  and the mean relation above is +0.3 $\pm$ 92.5 K.
If all giants, including the three anomalous stars, are compared, the values become +0.7 $\pm$ 98.9 K.

Surface gravity estimates ($\log g$) were obtained by direct comparison of $V$ magnitudes and $B-V$ colors 
to isochrones of \citet{VA06}, constructed for a scaled solar 
composition with [Fe/H] = -0.29 and an age of 1.85 Gyr, the same as the comparison presented 
in AT16. The isochrone's predicted magnitudes and colors were adjusted to match the cluster's reddening, 
$E(B-V)=0.058$ and apparent distance modulus, $12.75$ (AT16).  

Input estimates for the microturbulent velocity parameter were constructed using three prescriptions. For dwarfs 
within appropriate limits of $T_{\mathrm{eff}}$ and $\log g$, the formula of \citet{ED93} was used.  For giants with $\log g < 3.0$, 
a gravity-dependent formula, $v_t= 2.0 - 0.2$ $\log g$, was used. For subgiants and fainter red giants, we made use of a scheme developed by 
\citet{Br12} to analyze spectra of Kepler candidate G and K stars for which $T_{\mathrm{eff}}$ from colors and $\log g$ validated by asteroseismology were available. 
The formulation by \citet{Br12} produces $v_t$ estimations from a  $T_{\mathrm{eff}}$ and $\log g$ dependent formulation that   
meshes well with the \citet{ED93} and our previously used formula for giants if incremented by 0.2 km/sec. 

\onecolumngrid
\newpage
%\floattable
\begin{deluxetable}{ccccccccccccc}
\tablenum{3}
\tablecaption{Abundances for NGC 2506 Stars}
\tablewidth{0pt}
\tablecolumns{13}
%\tabletypesize\tiny
%\tablewidth{0pc}
%\setlength{\tabcolsep} {0.03in}
\tablehead{
\colhead{ID No.} & \colhead{$T_{NN}$} & \colhead{$[Fe/H]_{NN}$} & 
\colhead{$[Fe/H]_{RS}$} & \colhead{MAD} & \colhead{$N_{lines}$} & \colhead{$[Ca/H]$} &
\colhead{$[Si/H]$} & \colhead{$[Ni/H]$} & \colhead{$\sigma([Ni/H)$} & \colhead{$\rm{T}_{phot}$} &
\colhead{${\rm log} g$} & \colhead{$v_t$} }
\startdata
1108 & 5448 & -0.21 & -0.13 & 0.14 & 16 & 0.00 & -0.34 & -0.32 & 0.11 & 5372 & 3.23 & 1.35  \\
1112 & 4995 & -0.26 & -0.26 & 0.06 & 16 & .... & -0.28 & -0.34 & 0.04 & 5070 & 2.24 & 1.69  \\
1229 & 4821 & -0.34 & -0.34 & 0.06 & 16 & .... & -0.39 & -0.42 & 0.08 & 4856 & 2.32 & 1.65  \\
1301 & 5358 & -0.29 & -0.34 & 0.09 & 16 & -0.19 & -0.40 & -0.36 & 0.05 & 5249 & 3.23 & 1.35  \\
1320 & 4971 & -0.26 & -0.25 & 0.04 & 16 & .... & -0.29 & -0.39 & 0.03 & 5051 & 2.43 & 1.62  \\
1325 & 4967 & -0.30 & -0.29 & 0.07 & 16 & .... & -0.34 & -0.42 & 0.02 & 5070 & 2.34 & 1.65  \\
1340 & 5090 & -0.33 & -0.24 & 0.09 & 16 & 0.07 & -0.28 & -0.29 & 0.04 & 5308 & 3.18 & 1.37  \\
1377 & 5489 & -0.24 & -0.24 & 0.12 & 16 & -0.01 & -0.39 & -0.29 & 0.07 & 5421 & 3.29 & 1.36  \\
2212 & 4824 & -0.27 & -0.31 & 0.04 & 15 & .... & -0.37 & -0.46 & 0.03 & 4810 & 1.90 & 1.60  \\
2255 & 4933 & -0.31 & -0.26 & 0.06 & 16 & 0.05 & -0.37 & -0.42 & 0.06 & 5063 & 2.72 & 1.51  \\
2309 & 5061 & -0.25 & -0.29 & 0.04 & 16 & .... & -0.32 & -0.40 & 0.04 & 5065 & 2.31 & 1.66  \\
2329 & 5092 & -0.26 & -0.28 & 0.06 & 16 & 0.11 & -0.36 & -0.38 & 0.01 & 5106 & 2.35 & 1.65  \\
2351 & .... & .... & 0.14 & 0.29 & 10 & -0.06 & -0.07 & -0.02 & 0.10 & 7079 & 3.67 & 2.30  \\
2364 & 5289 & -0.24 & -0.14 & 0.07 & 16 & 0.28 & -0.26 & -0.27 & 0.02 & 5377 & 2.28 & 1.73  \\
2375 & 5058 & -0.16 & -0.30 & 0.07 & 16 & .... & -0.25 & -0.41 & 0.02 & 4990 & 2.60 & 1.55  \\
2380 & 5031 & -0.23 & -0.22 & 0.05 & 16 & .... & -0.37 & -0.36 & 0.06 & 5094 & 2.33 & 1.66  \\
2401 & 5015 & -0.26 & -0.25 & 0.08 & 16 & .... & -0.28 & -0.38 & 0.07 & 5111 & 2.37 & 1.65  \\
2402 & 4546 & -0.34 & -0.41 & 0.06 & 16 & .... & -0.27 & -0.50 & 0.05 & 4504 & 1.90 & 1.60  \\
3110 & 5383 & -0.25 & -0.24 & 0.15 & 15 & 0.02 & -0.34 & -0.31 & 0.05 & 5364 & 3.23 & 1.35  \\
3204 & 5235 & -0.29 & -0.21 & 0.08 & 16 & 0.04 & -0.34 & -0.35 & 0.07 & 5231 & 2.04 & 1.60  \\
      &      & 	    &	    &	   &	&      &       &       &       &     &      &       \\
3231 & 5022 & -0.28 & -0.27 & 0.04 & 16 & 0.04 & -0.30 & -0.37 & 0.04 & 5096 & 2.33 & 1.66  \\
3243 & 5467 & -0.21 & -0.15 & 0.09 & 16 & 0.05 & -0.41 & -0.29 & 0.04 & 5437 & 3.23 & 1.35  \\
3324 & 5045 & -0.24 & -0.24 & 0.05 & 16 & .... & -0.27 & -0.36 & 0.03 & 5089 & 2.43 & 1.62  \\
3356 & 5249 & -0.25 & -0.27 & 0.06 & 16 & 0.05 & -0.42 & -0.34 & 0.02 & 5200 & 2.99 & 1.42  \\
3359 & 5067 & -0.27 & -0.23 & 0.06 & 16 & .... & -0.24 & -0.38 & 0.04 & 5106 & 2.41 & 1.63  \\
3392 & 5096 & -0.27 & -0.27 & 0.07 & 16 & 0.06 & -0.35 & -0.37 & 0.05 & 5166 & 2.32 & 1.69  \\
4109 & 5364 & -0.27 & -0.24 & 0.07 & 16 & 0.07 & -0.30 & -0.41 & 0.06 & 5311 & 2.75 & 1.54  \\
4128 & 5255 & -0.27 & -0.26 & 0.08 & 17 & 0.05 & -0.39 & -0.48 & 0.07 & 5246 & 2.32 & 1.69  \\
4129 & .... & .... & -0.28 & 0.06 & 5 & -0.45 & -0.11 & -0.25 & 0.18 & 6314 & 3.54 & 1.83  \\
4138 & 5030 & -0.29 & -0.27 & 0.06 & 16 & .... & -0.27 & -0.34 & 0.03 & 5094 & 2.43 & 1.62  \\
4143 & 5058 & -0.22 & -0.31 & 0.08 & 16 & 0.05 & -0.34 & -0.43 & 0.03 & 5077 & 2.43 & 1.62  \\
4205 & 5082 & -0.24 & -0.21 & 0.09 & 16 & .... & -0.29 & -0.34 & 0.04 & 5094 & 2.42 & 1.63  \\
4223 & .... & .... & -0.30 & 0.06 & 12 & -0.18 & -0.28 & -0.43 & 0.13 & 5886 & 3.42 & 1.48  \\
4240 & 4992 & -0.28 & -0.29 & 0.07 & 15 & 0.03 & -0.35 & -0.43 & 0.02 & 5030 & 2.43 & 1.62  \\
4274 & 5168 & -0.26 & -0.27 & 0.08 & 16 & .... & -0.16 & -0.36 & 0.04 & 5035 & 2.73 & 1.50  \\
4372 & .... & .... & -0.40 & 0.18 & 6 & -0.54 & -1.14 & -0.85 & 0.18 & 6607 & 3.66 & 2.06  \\
4528 & 5534 & -0.19 & -0.22 & 0.09 & 16 & -0.05 & -0.33 & -0.38 & 0.06 & 5374 & 3.23 & 1.37  \\
5249 & 5526 & -0.34 & -0.55 & 0.17 & 16 & -0.50 & -0.50 & -0.65 & 0.09 & 5413 & 3.23 & 1.35  \\
5271 & 5059 & -0.26 & -0.25 & 0.04 & 16 & 0.10 & -0.36 & -0.38 & 0.02 & 5108 & 2.38 & 1.64  \\
5371 & 5401 & -0.23 & -0.18 & 0.07 & 16 & -0.05 & -0.46 & -0.30 & 0.04 & 5385 & 3.27 & 1.36  \\
      &      & 	    &	    &	   &	&      &       &       &       &     &      &       \\
7008 & 5423 & -0.28 & -0.34 & 0.10 & 16 & -0.05 & -0.33 & -0.33 & 0.09 & 5234 & 3.31 & 1.31  \\
7019 & 5234 & -0.26 & -0.21 & 0.06 & 16 & -0.06 & -0.39 & -0.35 & 0.09 & 5202 & 3.20 & 1.35  \\
7023 & 5090 & -0.25 & -0.29 & 0.06 & 16 & 0.04 & -0.35 & -0.47 & 0.04 & 5068 & 2.40 & 1.63  \\
7026 & 5344 & -0.23 & -0.34 & 0.06 & 16 & -0.17 & -0.31 & -0.46 & 0.06 & 5065 & 3.12 & 1.36  \\
7031 & 5611 & -0.48 & -0.43 & 0.17 & 11 & -0.34 & -0.34 & -0.48 & 0.07 & 5744 & 3.42 & 1.41  \\
7032 & .... & .... & -0.18 & 0.22 & 10 & -0.19 & -0.13 & -0.30 & 0.19 & 6161 & 3.65 & 1.64  \\
7036 & 5009 & -0.26 & -0.23 & 0.07 & 16 & -0.10 & -0.36 & -0.41 & 0.07 & 5051 & 2.42 & 1.62  \\
7041 & 5768 & 0.31 & 0.24 & 0.08 & 16 & .... & 0.28 & 0.31 & 0.06 & 5643 & 3.58 & 1.31  \\
7043 & 5283 & -0.34 & -0.27 & 0.07 & 15 & 0.02 & -0.41 & -0.22 & 0.03 & 5377 & 3.14 & 1.40  \\
7045 & 4736 & -0.13 & -0.31 & 0.04 & 15 & .... & -0.04 & -0.28 & 0.02 & 4741 & 2.52 & 1.57  \\
7048 & 4981 & -0.28 & -0.32 & 0.07 & 16 & .... & -0.30 & -0.43 & 0.03 & 5021 & 2.27 & 1.67  \\
7066 & .... & .... & -0.05 & 0.20 & 13 & -0.54 & -0.38 & -0.21 & 0.09 & 6761 & 3.70 & 2.30  \\
7069 & 5478 & 0.34 & 0.24 & 0.01 & 15 & .... & 0.27 & 0.08 & 0.09 & 5369 & 2.60 & 1.60  \\
7080 & 5377 & -0.24 & -0.21 & 0.10 & 15 & -0.05 & -0.32 & -0.38 & 0.10 & 5288 & 3.23 & 1.35  \\
7082 & 5712 & -0.31 & -0.27 & 0.16 & 13 & -0.22 & -0.31 & -0.38 & 0.19 & 5771 & 3.42 & 1.42  \\
7084 & 4750 & -0.29 & -0.34 & 0.04 & 15 & .... & -0.41 & -0.48 & 0.02 & 4787 & 2.33 & 1.64  \\
7085 & 4959 & -0.29 & -0.30 & 0.06 & 16 & .... & -0.21 & -0.43 & 0.04 & 4928 & 2.55 & 1.56  \\
7086 & 5247 & -0.37 & -0.17 & 0.11 & 16 & -0.13 & -0.23 & -0.28 & 0.10 & 5479 & 3.40 & 1.33  \\
7088 & .... & .... & -0.31 & 0.19 & 6 & -0.31 & -0.63 & -0.43 & 0.18 & 6329 & 3.54 & 1.83  \\
7093 & 4989 & 0.04 & 0.02 & 0.11 & 16 & .... & -0.02 & -0.15 & 0.07 & 5244 & 3.17 & 1.36  \\
      &      & 	    &	    &	   &	&      &       &       &       &     &      &       \\
7098 & 5099 & -0.30 & -0.34 & 0.04 & 15 & 0.01 & -0.29 & -0.49 & 0.05 & 5030 & 2.45 & 1.61  \\
7099 & 5055 & -0.25 & -0.29 & 0.05 & 16 & 0.04 & -0.31 & -0.43 & 0.06 & 5042 & 2.45 & 1.61  \\
7102 & 5154 & -0.22 & -0.27 & 0.06 & 16 & .... & -0.35 & -0.38 & 0.02 & 5070 & 2.39 & 1.63  \\
7106 & 4848 & -0.28 & -0.33 & 0.06 & 15 & .... & -0.35 & -0.50 & 0.06 & 4891 & 2.40 & 1.62  \\
7108 & 4959 & -0.29 & -0.32 & 0.07 & 16 & .... & -0.30 & -0.41 & 0.01 & 5007 & 2.63 & 1.54  \\
7114 & 5001 & -0.33 & -0.31 & 0.06 & 16 & -0.07 & -0.41 & -0.32 & 0.09 & 5080 & 3.03 & 1.39  \\
7117 & 5085 & -0.24 & -0.27 & 0.05 & 16 & .... & -0.33 & -0.38 & 0.04 & 5051 & 2.42 & 1.62  \\
7128 & 5228 & -0.25 & -0.24 & 0.07 & 15 & .... & -0.32 & -0.34 & 0.02 & 5115 & 3.03 & 1.40 \\
\enddata
\end{deluxetable}

\newpage
\twocolumngrid

A detailed discussion and application of the ROBOSPECT software \citep{WH13} and its optimization as applied to more than 
330 stars in NGC 6819 are supplied in \citet{LB15}. In short, our procedure consists of identification of the 
measurable spectroscopic absorption lines and calibration of atomic data, automated EW measurement 
using ROBOSPECT \citep{WH13}, atmospheric model construction using parameters ($T_{\mathrm{eff}}$, $\log g$, 
and $v_t$) as described above, and chemical abundance analysis using MOOG \citep{SN73}. 

Our ROBOSPECT line list is the same as used in the analysis of NGC 6819 \citep{LB15} and contains 
22 lines: 17 Fe I, 3 Ni I, 1 Si I, and 1 Ca I. These lines are unblended in the solar spectrum and have solar EW 
values in the range 10-150 m\AA, making them ideal for EW-based abundance analysis. For each line, we adopted 
atomic data (wavelength, excitation potential, and $\log gf$) contained in the VALD database \citep{KU99}. 
We then modified the retrieved $\log gf$ values such that our solar EW measures reproduced the solar elemental 
abundances given in the 2010 version of MOOG, using a \citet{KU92} model with 5770 K, 4.40, 1.14 km s$^{-1}$, 
and 0.00 for $T_{\mathrm{eff}}$, $\log g$, $v_t$, and [Fe/H], respectively.

Using our calibrated line list, 
ROBOSPECT then iteratively determines the continuum, noise, and line components of a spectrum and
reliably returns EWs that compare favorably with manual measurement of the lines. ROBOSPECT fits Gaussian line profiles, 
so we restrict our EW analysis to stars with projected 
rotation velocities $V_{rot} \leq 30$ km s$^{-1}$ in order to reduce systematic EW offsets due to non-Gaussian profiles. 
Additionally, we reject EWs greater than 150 m\AA\  to restrict our analysis to lines corresponding to the 
linear portion of the curve of growth. Finally, we discard EWs that are within 3$\sigma$ of the local noise level 
(calculated using a 6\AA\  window) to prevent introduction of spurious EW measurements into our analysis. 

Our sample of stars processed with ROBOSPECT began with 78 spectra drawn from the sample summarized in Table 1, choosing 
stars with M or MN designations and rotational velocities $\leq 30$ km/sec. Application of EW quality cuts left EW 
measures for 1022 Fe, 207 Ni, 52 Ca, and 70 Si lines out of a possible 1326, 234, 78, and 78, respectively. The majority of rejected Fe and Ni 
lines fell within our 3$\sigma$ significance threshold, while rejected Ca measurements generally had EWs $> 150$ m\AA. 
Our EW $< 150$ m\AA\  criterion is more stringent than the EW $< 200$ m\AA\  threshold used 
in \citet{LB15}, but adopting the more generous threshold does not change our abundance results. Nine of the 78 stars had fewer than 
five Fe lines remaining for analysis and were not included in the cluster average, as was one additional star with a $B-V = 1.77$, $T_{\mathrm{eff}}$ below 4000 K, and a spectrum dominated by molecular bands. Atmospheric parameters and EW abundances for each of the remaining 68 stars are provided in Table 3.

To translate between measured EW and [A/H], we first constructed a 1-D, plane-parallel model atmosphere for 
each star in our final ROBOSPECT sample using the \citet{KU92} model grid and the $T_{\mathrm{eff}}$, 
$\log g$, and $v_t$ values derived from our photometric observations. One star, 5270, has a temperature (7837 K)
greater than supported by the model atmosphere grid and was omitted from our sample. Our atmospheric models were then used in 
conjunction with our EW measurements as inputs to the \textit{abfind} MOOG routine, resulting in an [A/H] estimate for 
each of our measured EWs.  As shown in Figs. 3 and 4, respectively, there are no apparent trends of [Fe/H] with
$T_{\mathrm{eff}}$ or wavelength. There is, however, a serious decline in the number of turnoff stars with measurable abundances. 
The entire turnoff region is at $B-V$ $<$ 0.5 ($T_{\mathrm{eff}}$ $>$ 6400 K); only a handful of turnoff stars remain 
in the analysis set and these have large uncertainties. The reason for this is the declining line strength with increasing
$T_{\mathrm{eff}}$ at a metallicity less than one-half the solar value. Thus, the statistical averages are dominated by the
red giants. Among the red giants, the scatter in [Fe/H] (Fig. 4) is small, but three stars (7041, 7069, 7093) appear 
to have approximately solar abundance or higher, distinctly higher than the cluster mean, error bars included. Since only radial-velocity membership is available for all three stars, it is plausible that all three are field interlopers. This is especially likely for 7069, a star located well blueward of the FRG branch at $(B-V)$ = 0.85, $V$ = 13.56. The other two
stars fall along the subgiant branch and near the base of the giant branch. As a simple statistical check, if we plot the distribution of radial velocities for all red giants included in the study without {\it a priori} membership insight from proper motions and assume the field star histogram of Fig. 1, as defined 
by the red giants alone, is continuous across the radial-velocity distribution range of the cluster, the predicted number of field stars with a compatible radial velocity is found to be between 2 and 5, in excellent agreement with the three metal-rich stars identified here.

We find the cluster median iron abundance to be [Fe/H] $= -0.27 \pm 0.07$ dex, where the reported uncertainty 
is the median absolute deviation (MAD). This value was calculated by computing the median [Fe/H] for each star, 
and then using these values to compute the median/MAD [Fe/H] for our entire sample. This reported abundance value 
is robust under various calculation schemes; choosing to calculate [Fe/H] by taking the median of all Fe lines 
or by weighting each star's contribution to the overall [Fe/H] by the number of Fe lines measured produces 
nearly the same overall result. We adopt the median/MAD statistics here as 
our sample size is sufficiently large and these statistics are more robust against outliers, compared with using 
the mean/$\sigma$ statistics. In Table 3, we give the number of Fe lines measured, median and MAD for [Fe/H], as well as estimates for 
[Ni/H], [Ca/H], and [Si/H] for each star in our ROBOSPECT sample. The normalized MAD statistic, MADN $ = 1.48 \times$MAD, 
can be used to approximate the standard deviation of 
our [Fe/H] estimate. Using the MADN, we find the iron abundance for NGC 2506 to be [Fe/H] $= -0.27 \pm 0.013$ dex (s.e.m.). 
This result is the same as that generated by ANNA and only slightly higher than our photometric abundance estimate 
of [Fe/H] = -0.32 $\pm$ 0.03 (AT16).

Abundances and atmospheric parameters for each of the 65 stars with ROBOSPECT-derived abundances are reported in Table 3. For most stars we present 
the median abundance from
5 to 17 Fe lines along with the MAD for that star. The [Ni/H] abundance estimates are not based on median values for each star, since there are at most three Ni lines in our spectral region.  We converted the logarithmic value for each abundance measurement to numerical values before averaging and computing the standard deviation among measurements, then converting the Ni abundance to a logarithmic [Ni/H] value.  The quoted error represents the effect on [Ni/H] of a standard deviation added to the numerical Ni abundance value.
For the sample presented in Table 3, the cluster median value of [Ca/H] is $-0.01 \pm 0.06$, [Si/H] $= -0.33 \pm 0.05$, and [Ni/H] $= -0.38 \pm 0.05$. For the Ca and Si estimates, the quoted error is the median absolute deviation of all of the single-star abundance estimates relative to the cluster median value; for Ni, the quoted error is the median value of each star's estimated error, as described above.  

To get a handle on the impact of possible errors in the input parameters, the abundances were redetermined under the
assumption that $T_{\mathrm{eff}}$ was altered by $\pm$ 100 K, $\log g$ by $\pm$ 0.25 and $v_t$ by $\pm$ 0.25 km s$^{-1}$, an approach followed in our analysis of NGC 6819 \cite{LB15}. Our results are similar to those presented therein:  the effect of altering $T_{\mathrm{eff}}$, $\log g$ and $v_t$ by the amounts specified above is to increment the abundance [Fe/H] by $\pm$0.06, 0.00 and $\mp$0.06, respectively, for stars bluer than (B-V) = 0.8. Increments for redder stars are similar: $\pm$0.06, $\mp$0.03 and $\mp$0.06, respectively.  

\begin{figure}
\figurenum{3}
\plotone{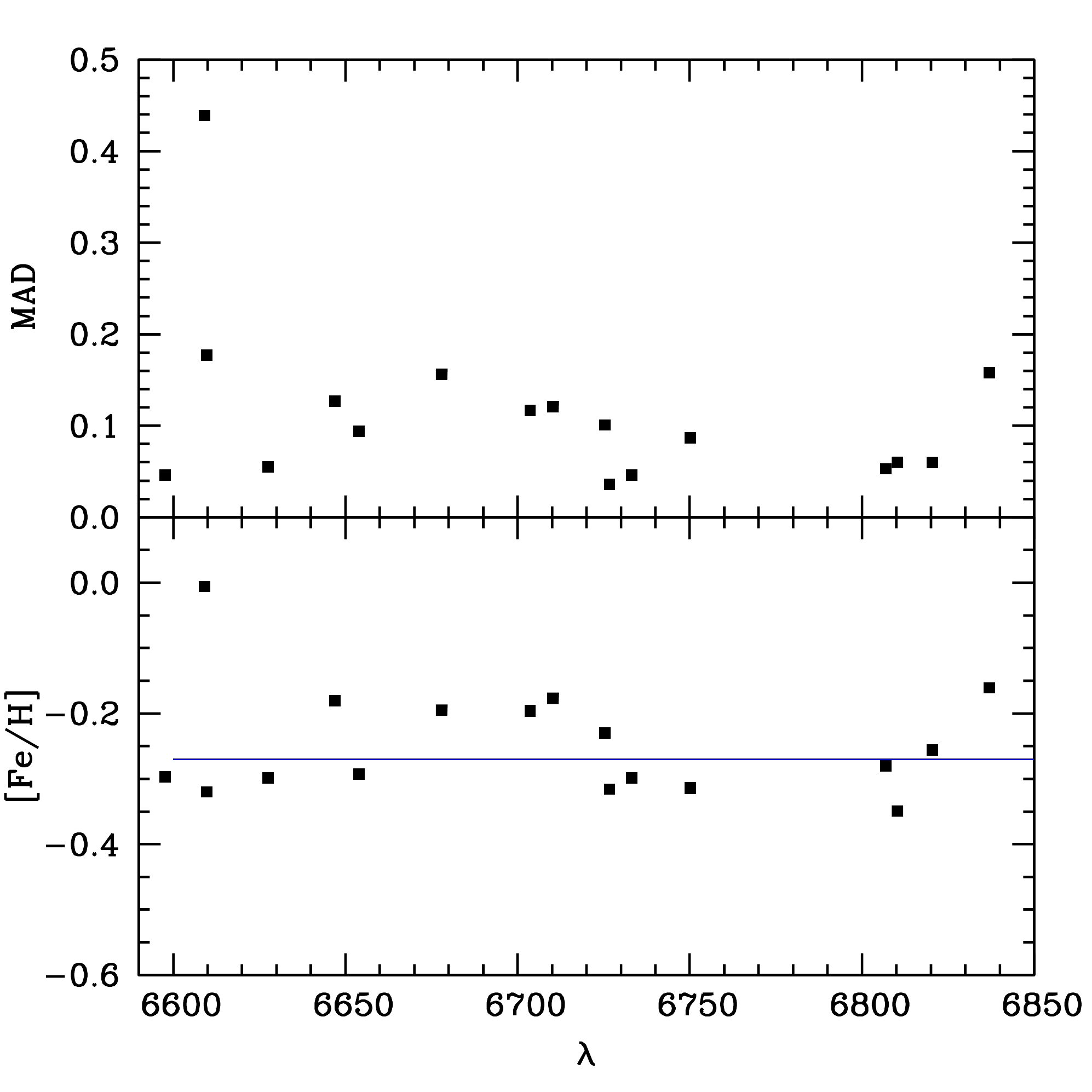}
\caption{For each of the 17 Fe lines the abundance [Fe/H] is shown as well as the MAD (median absolute deviation) for our spectroscopic sample. The blue horizontal line shows the cluster sample median value of [Fe/H] $=n-0.27$.}
\end{figure}

\begin{figure}
\figurenum{4}
\plotone{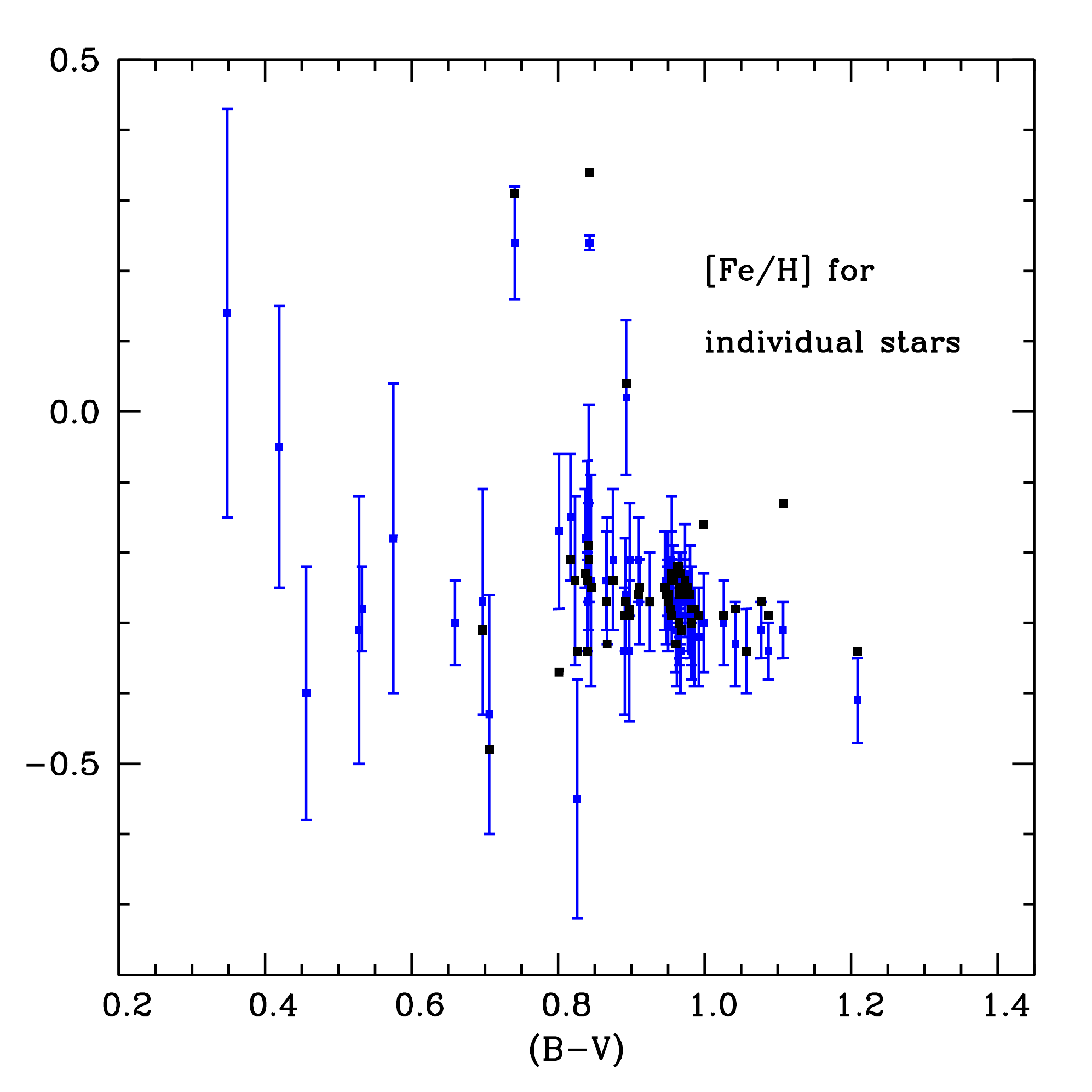}
\caption{For each star, the median [Fe/H] from ROBOSPECT analysis is shown as a function of (B-V) color using blue symbols; the error bars indicate the size of the MAD statistic for that star.  Black points designate [Fe/H] estimates from the ANNA neural network analysis.  }
\end{figure}

\subsection{Comparison to Previous Work}
There have been three studies of NGC 2506 tied to high dispersion spectroscopy \citep{RE12, MI11, CA16}, all using red giant 
samples dominated by clump stars. For the three studies sampled, the overlap with the current investigation is 3, 2, and 4 stars; for
\citet{CA16}, star 3265 was excluded due to its peculiar nature, probably a signature of binarity. In each case, we can include
two $T_{\mathrm{eff}}$ ($T_{\mathrm{NN}}$ and $T_{\mathrm{ph}}$) and two [Fe/H] ([Fe/H]$_{NN}$ and [Fe/H]$_{RS}$) comparisons.
The [Fe/H]$_{RS}$ refers to metallicity based upon analysis using the $T_{\mathrm{ph}}$ and equivalent widths measured by ROBOSPECT.
In all three $T_{\mathrm{eff}}$ comparisons, the ANNA scale and the color-based scale are systematically hotter than the published values, with
the color-based $T_{\mathrm{eff}}$ always slightly hotter than the ANNA scale. The offsets for the two scales are +24 $\pm$ 96 and +66 $\pm$ 63,
+6 $\pm$ 22 and +51 $\pm$ 5, +19 $\pm$ 28 and +63 $\pm$ 37, for \citet{RE12, MI11, CA16}, respectively. The analogous offsets for [Fe/H] from ANNA and
ROBOSPECT are -0.06 $\pm$ 0.03 and -0.07 $\pm$ 0.05, -0.03 $\pm$ 0.07 and -0.01 $\pm$ 0.04, and 0.03 $\pm$ 0.04 and 0.04 $\pm$ 0.04, respectively.
Since the adopted $T_{\mathrm{eff}}$ in the current investigation is hotter in all cases, transfer of the current data to the published systems
will lower our [Fe/H] in all cases, by typically 0.01 to 0.02 dex for $T_{\mathrm{NN}}$ and 0.03 to 0.04 dex for $T_{\mathrm{ph}}$.
Within the uncertainties of a small sample, this implies that we are on the same metallicity scale as \citet{CA16}.

For the elements other than Fe, comparisons can be made to the discussions of \citet{MI11, RE12}. \citet{MI11} find [Ca/Fe], [Si/Fe], and [Ni/Fe] =
-0.06 $\pm$ 0.06, 0.01 $\pm$ 0.07, and -0.08 $\pm$ 0.06, respectively, while \citet{RE12} get 0.10 $\pm$ 0.05, +0.04 $\pm$ 0.04, and -0.08 $\pm$ 0.04,
respectively. Our [Ca/Fe] = +0.28 $\pm$ 0.07 is based upon one generally strong line of Ca and is inconsistent with both the published spectroscopic and photometric, via the CaII-based $hk$ index, abundances which indicate an effectively solar ratio within the uncertainties. It should be
regarded as suspect. By contrast, we find [Si/Fe] = -0.06 $\pm$ 0.05 and [Ni/Fe] = -0.11 $\pm$ 0.05. In all three studies, 
Si is differentially more metal-rich than Ni by an average of +0.09 $\pm$ 0.04 dex. The dominant source of uncertainty in our ratios lies with the Si abundance, tied to a single line of modest equivalent width in each star.

\section{Lithium}

\subsection{Abundance Estimation}

We used the SPLOT utility within IRAF to measure equivalent widths (EW) and Gaussian full-widths of the Li 6707.8 \AA\ line for all 287 stars.  A reasonably close and relatively line-free region between 6680 \AA\ and 6690 \AA\ was examined to estimate the signal-to-noise ratio per pixel, also 
estimated using SPLOT. 
Our analysis of the EW measurements employs a computational scheme that 
first numerically removes the contribution to the Li line EW produced by the nearby Fe I line 
at 6707.45 \AA, assuming a cluster metallicity of [Fe/H] = -0.30 and a temperature estimate for the star.  
For the latter input, we used the same color-temperature scheme as described earlier. The computational 
scheme then interpolates within a model-atmosphere-generated grid of EWs and temperatures to 
estimate a Li abundance for each star, a scheme developed by \citet{ST03} and employed by \citet{SD04}.  
A star is only considered to have a detected Li abundance if the corrected EW exceeds three times the estimated 
error in the EW, itself a function of the measured line width and SNR for the spectrum \citep{DE93, DP93}.  
For stars with EW below this criterion, the Li abundance can only be characterized as an upper limit and no abundance error is estimated.
The abundance error estimate for detected lines is primarily dependent on the error in the EW.  We did test the sensitivity of the computed Li abundance 
to increments of $+100$ K in $T_{\mathrm{eff}}$. A higher $T_{\mathrm{eff}}$ results in higher Li abundance but by widely different amounts for different classes of stars;
for warmer dwarfs, the abundance increment is $\leq 0.1$, rising to $0.15$ for cooler subgiants and high sensitivity ($\geq 0.3$) for stars on the red giant branch.

A more subtle concern for the coolest stars arises from the presence of CN lines in the region of the Li line. From synthetic spectra, we find that at the metallicity of NGC 2506 and plausible levels of CN enhancement, the molecular lines have no impact on the measured EW. For stars of solar metallicity and higher, CN-enhanced giants can be affected. 

Of the 175 stars classed as single-star members, Li detections or upper limits were possible for all but 30. 
Of the 30 stars, one is 4402, the coolest red giant member of our sample with $B-V$ = 1.76.  
The confusion and complexity of features near the Li line made any attempt at a Li abundance estimate for this star impossible. 
This internal photoelectric standard \citep{MC81} lies outside the astrometric survey \citep{CA81} and, like 2402, has
generally been ignored in spectroscopic surveys of the red giant branch.
Its location in the CMD, however, places this star closest to the tip of the red giant branch and a
potential candidate for evidence of Li-enrichment \citep{CA16}; observation at a higher resolution could
prove informative.
 
The 29 remaining stars fall within a well-defined group: rapidly rotating stars ($V_{rot} >$ 30 km s$^{-1}$) at 
the CMD turnoff. The average measured rotation speed for the excluded stars is 54.2 km s$^{-1}$, broadening the Li line,
if one exists, to a level where, when combined with the statistical noise, any attempt at directly measuring or even 
placing a constraint on the Li value proved implausible. 
The A(Li) values with non-zero uncertainties for stars with measureable Li and zero for stars with upper limits are
presented in Table 4. 

\onecolumngrid
\newpage
%\floattable
\begin{deluxetable}{cccccccc}
\tablenum{4}
\tablecaption{Lithium Abundances for NGC 2506 Stars}
\tablewidth{0pt}
\tablecolumns{8}
%\tabletypesize\tiny
%\tablewidth{0pc}
%\setlength{\tabcolsep} {0.03in}
\tablehead{
\colhead{ID No.} & \colhead{$\rm{T}_{eff} $} & \colhead{EW(Li)} &
\colhead{SNR} & \colhead{FWHM} & \colhead{A(Li)} &
\colhead{$\sigma_{ALi}$} & \colhead{Code} }
\startdata 
     &      &       &  	&	&	& 	& \\
1108 & 5372 & 21.2 & 124 & 0.46 & 0.70 & 0.00 & 2 \\
1112 & 5070 & 16.6 & 129 & 0.80 & 0.55 & 0.00 & 3 \\
1127 & 6909 & 5.0 & 132 & 0.70 & 2.50 & 0.00 & 1 \\
1134 & 6950 & 30.5 & 158 & 1.60 & 2.94 & 0.09 & 0 \\
1211 & 7004 & 30.0 & 195 & 1.01 & 2.96 & 0.06 & 0 \\
1229 & 4856 & 53.7 & 124 & 0.67 & 1.11 & 0.06 & 0 \\
1301 & 5249 & 45.9 & 118 & 0.89 & 1.56 & 0.08 & 0 \\
1302 & 6950 & 23.0 & 117 & 0.90 & 2.80 & 0.12 & 0 \\
1305 & 7009 & 29.9 & 187 & 1.03 & 2.97 & 0.06 & 0 \\
1320 & 5051 & 13.6 & 143 & 0.64 & 0.55 & 0.00 & 3 \\
1325 & 5070 & 10.9 & 138 & 0.53 & 0.55 & 0.00 & 3 \\
1331 & 6652 & 45.4 & 155 & 0.94 & 2.94 & 0.05 & 0 \\
1340 & 5308 & 27.0 & 179 & 0.67 & 1.24 & 0.10 & 0 \\
1350 & 6936 & 56.2 & 170 & 1.94 & 3.25 & 0.05 & 0 \\
1354 & 6968 & 41.0 & 120 & 0.98 & 3.10 & 0.07 & 0 \\
1358 & 6852 & 34.6 & 198 & 2.04 & 2.94 & 0.07 & 0 \\
1377 & 5421 & 37.1 & 180 & 0.65 & 1.64 & 0.06 & 0 \\
1379 & 6959 & 27.8 & 175 & 1.10 & 2.90 & 0.07 & 0 \\
1384 & 6735 & 67.4 & 236 & 2.42 & 3.21 & 0.04 & 0 \\
1390 & 6936 & 5.0 & 140 & 0.70 & 2.49 & 0.00 & 1 \\
1394 & 6702 & 71.4 & 130 & 1.50 & 3.23 & 0.05 & 0 \\
2102 & 6847 & 44.0 & 188 & 2.30 & 3.06 & 0.07 & 0 \\
2140 & 6856 & 30.2 & 170 & 0.82 & 2.87 & 0.06 & 0 \\
2212 & 4810 & 26.1 & 96 & 0.68 & 0.40 & 0.00 & 3 \\
2217 & 6635 & 79.2 & 160 & 1.66 & 3.24 & 0.04 & 0 \\
2255 & 5063 & 39.4 & 147 & 0.72 & 1.18 & 0.08 & 0 \\
2306 & 6963 & 49.0 & 113 & 1.31 & 3.19 & 0.08 & 0 \\
2309 & 5065 & 9.4 & 145 & 0.58 & 0.55 & 0.00 & 3 \\
2311 & 6640 & 66.5 & 156 & 1.59 & 3.14 & 0.05 & 0 \\
2325 & 6968 & 5.0 & 100 & 0.70 & 2.67 & 0.00 & 1 \\
     &      &       &  	&	&	& 	& \\
2328 & 6860 & 70.2 & 199 & 2.90 & 3.32 & 0.05 & 0 \\
2329 & 5106 & 12.4 & 156 & 0.56 & 0.55 & 0.00 & 3 \\
2332 & 6869 & 8.2 & 148 & 0.97 & 2.49 & 0.00 & 1 \\
2343 & 6945 & 32.7 & 94 & 0.70 & 2.97 & 0.09 & 0 \\
2347 & 6865 & 36.0 & 229 & 1.62 & 2.96 & 0.05 & 0 \\
2351 & 7079 & 52.2 & 250 & 1.50 & 3.30 & 0.03 & 0 \\
2364 & 5377 & 20.5 & 169 & 0.85 & 0.70 & 0.00 & 2 \\
2371 & 6927 & 32.5 & 174 & 1.02 & 2.95 & 0.06 & 0 \\
2373 & 6963 & 23.6 & 130 & 1.03 & 2.82 & 0.11 & 0 \\
2375 & 4990 & 52.2 & 146 & 0.77 & 1.29 & 0.06 & 0 \\
2380 & 5094 & 16.4 & 175 & 0.81 & 0.55 & 0.00 & 3 \\
2386 & 6945 & 39.9 & 206 & 1.25 & 3.07 & 0.05 & 0 \\
2394 & 7000 & 37.8 & 122 & 1.29 & 3.08 & 0.09 & 0 \\
2401 & 5111 & 17.4 & 163 & 0.75 & 0.55 & 0.00 & 3 \\
2402 & 4505 & 60.4 & 94 & 0.71 & 0.59 & 0.10 & 0 \\
3110 & 5364 & 25.0 & 159 & 0.68 & 1.26 & 0.13 & 0 \\
3112 & 7046 & 25.0 & 103 & 1.03 & 2.90 & 0.13 & 0 \\
3135 & 6891 & 32.8 & 115 & 1.10 & 2.94 & 0.10 & 0 \\
3158 & 6765 & 65.0 & 115 & 0.67 & 3.21 & 0.04 & 0 \\
3204 & 5231 & 9.2 & 145 & 0.66 & 0.70 & 0.00 & 3 \\
3206 & 6740 & 45.4 & 199 & 2.25 & 3.00 & 0.06 & 0 \\
3231 & 5096 & 12.4 & 142 & 0.59 & 0.55 & 0.00 & 3 \\
3263 & 6896 & 26.8 & 176 & 1.20 & 2.84 & 0.08 & 0 \\
3324 & 5089 & 15.6 & 146 & 0.62 & 0.55 & 0.00 & 3 \\
3344 & 6991 & 44.6 & 134 & 1.10 & 3.16 & 0.06 & 0 \\
3350 & 6918 & 52.7 & 174 & 1.59 & 3.20 & 0.05 & 0 \\
3356 & 5200 & 36.2 & 160 & 0.75 & 1.32 & 0.08 & 0 \\
3359 & 5106 & 12.5 & 164 & 0.61 & 0.55 & 0.00 & 3 \\
3360 & 6778 & 76.0 & 194 & 1.72 & 3.31 & 0.04 & 0 \\
3367 & 6968 & 24.2 & 160 & 1.00 & 2.84 & 0.09 & 0 \\
     &      &       &  	&	&	& 	& \\
3373 & 6986 & 30.6 & 109 & 0.89 & 2.96 & 0.10 & 0 \\
3394 & 6927 & 5.0 & 155 & 0.70 & 2.43 & 0.00 & 1 \\
4105 & 7051 & 23.4 & 138 & 1.33 & 2.87 & 0.12 & 0 \\
4106 & 6887 & 5.0 & 157 & 0.70 & 2.40 & 0.00 & 1 \\
4109 & 5311 & 31.8 & 180 & 0.68 & 1.38 & 0.08 & 0 \\
4128 & 5246 & 61.4 & 177 & 0.67 & 1.75 & 0.03 & 0 \\
4129 & 6314 & 64.6 & 118 & 0.98 & 2.88 & 0.05 & 0 \\
4138 & 5094 & 20.6 & 172 & 0.82 & 0.55 & 0.00 & 2 \\
4145 & 6959 & 21.8 & 120 & 0.88 & 2.78 & 0.12 & 0 \\
4205 & 5094 & 19.7 & 162 & 0.90 & 0.55 & 0.00 & 2 \\
4216 & 6731 & 74.8 & 195 & 1.56 & 3.27 & 0.04 & 0 \\
4218 & 7327 & 30.6 & 161 & 0.98 & 3.18 & 0.07 & 0 \\
4223 & 5886 & 60.8 & 200 & 0.77 & 2.48 & 0.03 & 0 \\
4240 & 5030 & 13.5 & 119 & 0.70 & 0.55 & 0.00 & 3 \\
4274 & 5035 & 37.4 & 160 & 0.92 & 1.09 & 0.09 & 0 \\
4331 & 6883 & 61.4 & 198 & 2.78 & 3.26 & 0.05 & 0 \\
4337 & 6575 & 84.4 & 170 & 1.88 & 3.24 & 0.04 & 0 \\
4353 & 6968 & 26.8 & 145 & 0.99 & 2.89 & 0.09 & 0 \\
4361 & 6635 & 43.2 & 213 & 1.88 & 2.90 & 0.05 & 0 \\
4372 & 6607 & 53.1 & 188 & 1.13 & 2.99 & 0.04 & 0 \\
4528 & 5374 & 38.8 & 160 & 0.69 & 1.62 & 0.06 & 0 \\
5196 & 6977 & 32.0 & 148 & 1.35 & 2.98 & 0.08 & 0 \\
5214 & 6453 & 69.5 & 208 & 1.50 & 3.03 & 0.03 & 0 \\
5233 & 6909 & 58.2 & 190 & 1.61 & 3.25 & 0.04 & 0 \\
5249 & 5413 & 28.8 & 145 & 0.77 & 1.46 & 0.10 & 0 \\
5270 & 7600 & 5.0 & 230 & 0.70 & 2.69 & 0.00 & 1 \\
5271 & 5108 & 13.5 & 144 & 0.65 & 0.55 & 0.00 & 3 \\
5371 & 5385 & 5.0 & 135 & 0.70 & 0.70 & 0.00 & 3 \\
7008 & 5234 & 19.3 & 138 & 0.97 & 0.70 & 0.00 & 3 \\
7015 & 6847 & 41.4 & 180 & 1.58 & 3.02 & 0.06 & 0 \\
   &      &       &  	&	&	& 	& \\
7019 & 5202 & 33.7 & 159 & 0.64 & 1.27 & 0.07 & 0 \\
7022 & 6852 & 49.0 & 200 & 3.82 & 3.11 & 0.07 & 0 \\
7023 & 5068 & 17.3 & 159 & 0.63 & 0.55 & 0.00 & 3 \\
7026 & 5065 & 38.8 & 145 & 0.67 & 1.17 & 0.08 & 0 \\
7031 & 5744 & 76.0 & 200 & 0.89 & 2.47 & 0.03 & 0 \\
7032 & 6161 & 74.2 & 193 & 0.99 & 2.84 & 0.03 & 0 \\
7035 & 6472 & 71.6 & 165 & 0.99 & 3.07 & 0.03 & 0 \\
7036 & 5051 & 18.3 & 129 & 0.59 & 0.55 & 0.00 & 3 \\
7038 & 6786 & 67.8 & 165 & 2.44 & 3.25 & 0.06 & 0 \\
7041 & 5643 & 66.2 & 129 & 0.77 & 2.27 & 0.04 & 0 \\
7042 & 6765 & 55.4 & 185 & 1.46 & 3.12 & 0.04 & 0 \\
7043 & 5377 & 46.4 & 124 & 0.78 & 1.74 & 0.07 & 0 \\
7045 & 4741 & 21.9 & 89 & 0.63 & 0.20 & 0.00 & 3 \\
7047 & 6968 & 36.0 & 140 & 1.12 & 3.03 & 0.07 & 0 \\
7048 & 5021 & 17.9 & 150 & 0.84 & 0.55 & 0.00 & 3 \\
7053 & 6896 & 46.7 & 177 & 1.21 & 3.12 & 0.05 & 0 \\
7058 & 6932 & 30.3 & 151 & 1.10 & 2.92 & 0.08 & 0 \\
7059 & 6923 & 65.6 & 173 & 3.24 & 3.32 & 0.06 & 0 \\
7061 & 6689 & 6.0 & 198 & 0.70 & 2.16 & 0.00 & 1 \\
7062 & 6830 & 5.0 & 230 & 0.70 & 2.18 & 0.00 & 1 \\
7063 & 6744 & 51.0 & 195 & 2.01 & 3.06 & 0.05 & 0 \\
7064 & 6941 & 40.5 & 157 & 1.16 & 3.07 & 0.06 & 0 \\
7066 & 6761 & 9.0 & 261 & 0.50 & 2.23 & 0.10 & 0 \\
7069 & 5369 & 30.7 & 105 & 0.75 & 1.44 & 0.14 & 0 \\
7071 & 7107 & 34.6 & 125 & 1.80 & 3.10 & 0.11 & 0 \\
7076 & 6808 & 48.2 & 125 & 1.55 & 3.08 & 0.08 & 0 \\
7080 & 5288 & 26.4 & 186 & 0.64 & 1.19 & 0.10 & 0 \\
7081 & 6905 & 24.4 & 164 & 0.86 & 2.80 & 0.08 & 0 \\
7082 & 5771 & 54.7 & 189 & 0.68 & 2.29 & 0.03 & 0 \\
7084 & 4787 & 28.3 & 120 & 0.64 & 0.20 & 0.00 & 2 \\
     &      &       &  	&	&	& 	& \\
7085 & 4928 & 55.6 & 168 & 0.77 & 1.25 & 0.05 & 0 \\
7086 & 5479 & 51.9 & 140 & 0.95 & 1.93 & 0.06 & 0 \\
7088 & 6329 & 73.0 & 156 & 0.87 & 2.97 & 0.03 & 0 \\
7089 & 6673 & 66.0 & 242 & 2.71 & 3.16 & 0.04 & 0 \\
7090 & 6950 & 45.4 & 162 & 1.11 & 3.14 & 0.05 & 0 \\
7093 & 5244 & 25.4 & 124 & 0.98 & 0.70 & 0.00 & 2 \\
7094 & 6778 & 49.0 & 193 & 1.18 & 3.06 & 0.04 & 0 \\
7097 & 6619 & 38.8 & 201 & 1.82 & 2.84 & 0.06 & 0 \\
7098 & 5030 & 10.2 & 176 & 0.69 & 0.55 & 0.00 & 3 \\
7099 & 5042 & 20.4 & 146 & 0.82 & 0.55 & 0.00 & 2 \\
7102 & 5070 & 17.0 & 156 & 0.78 & 0.55 & 0.00 & 3 \\
7104 & 6757 & 46.6 & 192 & 1.38 & 3.02 & 0.05 & 0 \\
7105 & 6918 & 67.7 & 143 & 1.73 & 3.34 & 0.05 & 0 \\
7106 & 4891 & 55.0 & 127 & 0.69 & 1.19 & 0.06 & 0 \\
7108 & 5007 & 41.8 & 155 & 0.73 & 1.15 & 0.07 & 0 \\
7109 & 6900 & 24.6 & 132 & 1.42 & 2.80 & 0.12 & 0 \\
7110 & 6923 & 5.0 & 118 & 0.70 & 2.56 & 0.00 & 1 \\
7111 & 6972 & 16.1 & 191 & 1.09 & 2.65 & 0.11 & 0 \\
7114 & 5080 & 12.8 & 180 & 0.64 & 0.55 & 0.00 & 3 \\
7115 & 7102 & 32.0 & 116 & 0.61 & 3.06 & 0.07 & 0 \\
7117 & 5051 & 22.8 & 165 & 0.79 & 0.55 & 0.00 & 2 \\
7123 & 6727 & 58.2 & 232 & 0.91 & 3.12 & 0.03 & 0 \\
7125 & 6865 & 24.8 & 155 & 1.20 & 2.78 & 0.09 & 0 \\
7128 & 5115 & 39.7 & 180 & 0.76 & 1.27 & 0.06 & 0 \\
7132 & 6843 & 62.8 & 169 & 2.18 & 3.24 & 0.05 & 0 \\
\enddata
\tablecomments{Lithium abundance values with non-zero errors (code 0) are detections; stars with abundance codes 1, 2 or 3 are upper limits constrained by SNR, the temperature grid, or both, respectively.}
\end{deluxetable}

\newpage
\twocolumngrid
\subsection{Evolution in the CMD}

For the remaining 145 stars, Fig. 5 shows the CMD for stars with Li determinations; open black circles 
are detections while open red triangles designate upper limits. Blue symbols are detections for stars classed
as subgiants, allowing one to uniquely distinguish these stars from turnoff stars of the same B-V but 
at fainter magnitudes and from stars classed as FRGs in later figures. Filled symbols identify the three stars
which are spectroscopically determined to be metal-rich and therefore likely field stars, two of which 
have Li detections while the third has only an upper limit. Of the 41 stars with only upper limits to A(Li), 
10 are located at the turnoff (defined as $B-V  <$ 0.5, irrespective of $V$) and 31 populate the giant 
region ($B-V >$ 0.8). Among the redder stars, the sample splits into two distinct groups, stars at the 
base of the vertical turnup of the red giant branch and the red giant clump stars. With the exception 
of one anomalously blue clump star (4128), every giant located within the CMD region associated with the 
red clump has, at best, an upper limit for A(Li). It is crucial to note that this is not simply an issue 
of the clump stars exhibiting lower B-V and higher than average $T_{\mathrm{eff}}$ compared to FRGs at 
the same magnitude level and therefore weaker lines. The pattern among FRGs at the same $T_{\mathrm{eff}}$ 
but positioned 0.5 mag fainter than the red clump clearly demonstrates that if the clump stars had
A(Li) similar to the FRGs, it should be detectable. 

\begin{figure}
\figurenum{5}
\plotone{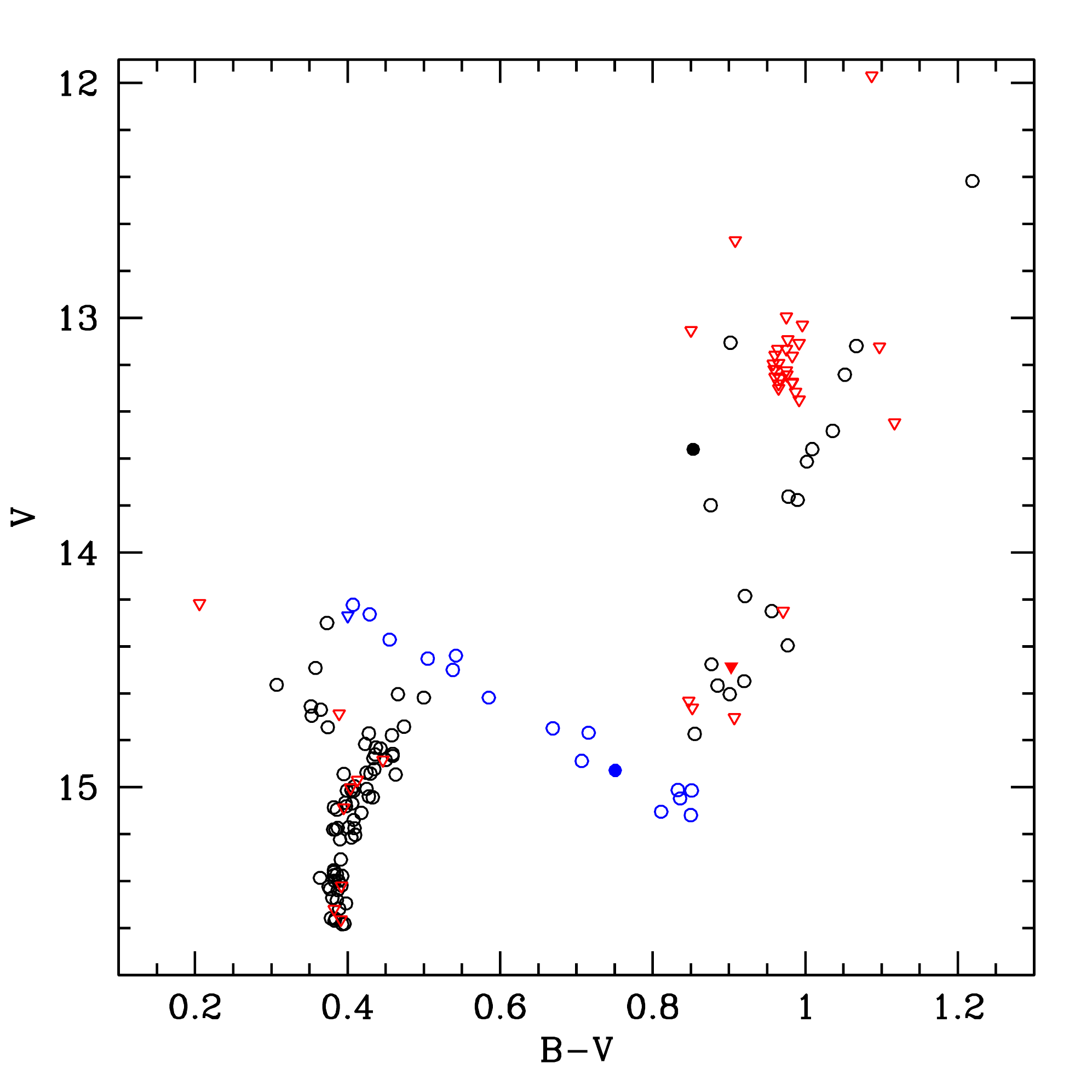}
\caption{CMD for single-star members with detectable Li (open black circles) and upper limits to A(Li) (open red triangles).
Blues stars identify the location of purported members of the subgiant branch. Filled symbols identify the three stars 
with estimated metallicities significantly higher than the cluster.}
\end{figure}

To probe the empirical trends among A(Li), we show in Fig. 6 the variation in A(Li) as a function of $V$. Symbols 
have the same meaning as in Fig. 5. The pattern with decreasing $V$ is a clear reflection of the evolution from
the main sequence to the giant branch, with the apparent reversal between $V$ = 15.2 and 14.6 caused by the shape 
of the subgiant branch. The stars at the turnoff, irrespective of magnitude, show an approximately constant mean 
A(Li) near 3.05 with a range approaching $\pm$0.30 dex. A(Li) steadily declines across the subgiant branch, up 
to the base of the FRG branch, with every subgiant having detectable Li, rather than just an upper limit. Between 
the base of the FRG at $V$ = 14.5 and the approximate luminosity level of the red giant clump at $V$ = 13.0, 
A(Li) among the stars with detectable Li remains effectively constant near A(Li) = 1.25. Only one star (2402) 
above the clump level has measurable Li, with A(Li) just below 0.6. As a newly identified probable member of the 
cluster, this star plays a unique role in redefining the location of the luminous end of the FRG branch.
Previous discussions \citep{CA04, MI11, RE12} have assumed that the FRG branch above the clump passed through 
the positions occupied by stars 2212 ($(B-V)$ = 1.09, $V$ = 11.97) and 2122 (1.10, 11.7)(not included in the figure). 
 The former star, included in this analysis, only has an upper limit to A(Li), as expected if it is a post-He-flash 
star. (But see the discussion on star 4128, below, on the possibility of Li production by post-He-flash stars.) 
If 2402 is truly a cluster member defining the FRG branch, then 2212 and 2122 lie almost a magnitude above 
this extension and cannot be normal, single stars in this same phase of evolution, i.e. they cannot be FRGs.
 
The fact that Li is detectable in 2402 at a level comparable to that found as an upper limit among the bluer and fainter 
clump stars is due in part to its having the coolest $T_{\mathrm{eff}}$ of any star with a measurable spectrum 
included in the membership sample. However, if its proposed evolutionary state is correct, it could be evidence 
that A(Li) does decline among FRGs beyond the level of the clump and prior to the He-flash, a point returned to below.
 
One star (4128) near $V$ = 13.1 has A(Li) = 1.75, higher than the average of 1.25 for FRG stars over a range in 
luminosity. It is located in Fig. 5 blueward of the red giant clump, leading to potential classification as a 
Li-rich giant since it sits just above the relatively fluid boundary used to define such stars 
\citep[e.g.,][]{KU11, AG14, CS16, TA17}. From isochrone comparisons \citep{VA06}, the initial mass of the stars 
populating the giant branch in NGC 2506 is between 1.6 and 1.7 $M_{\sun}$.  A(Li) $\sim$ 1.5 is what standard 
models predict for a star of 1.5 $M_{\sun}$ and solar abundance after the first dredge-up \citep{PA11}, but not 
for post He-flash red giants. Empirically, however, the star has A(Li) $\sim$ 0.5 dex above the observed value for the 
FRGs, irrespective of the predictions of standard models, which may be a signature of actual Li production in 
a star in the post-He-core-flash phase \citep{AG14, MO14}.  No other star on the giant branch, above or below the clump, exhibits Li sufficiently high enough to be tagged as a potential Li-rich star, though this is not unexpected given the 
often quoted 1$\%$ rate of detection for these stars in the general population of red giants.

\begin{figure}
\figurenum{6}
\plotone{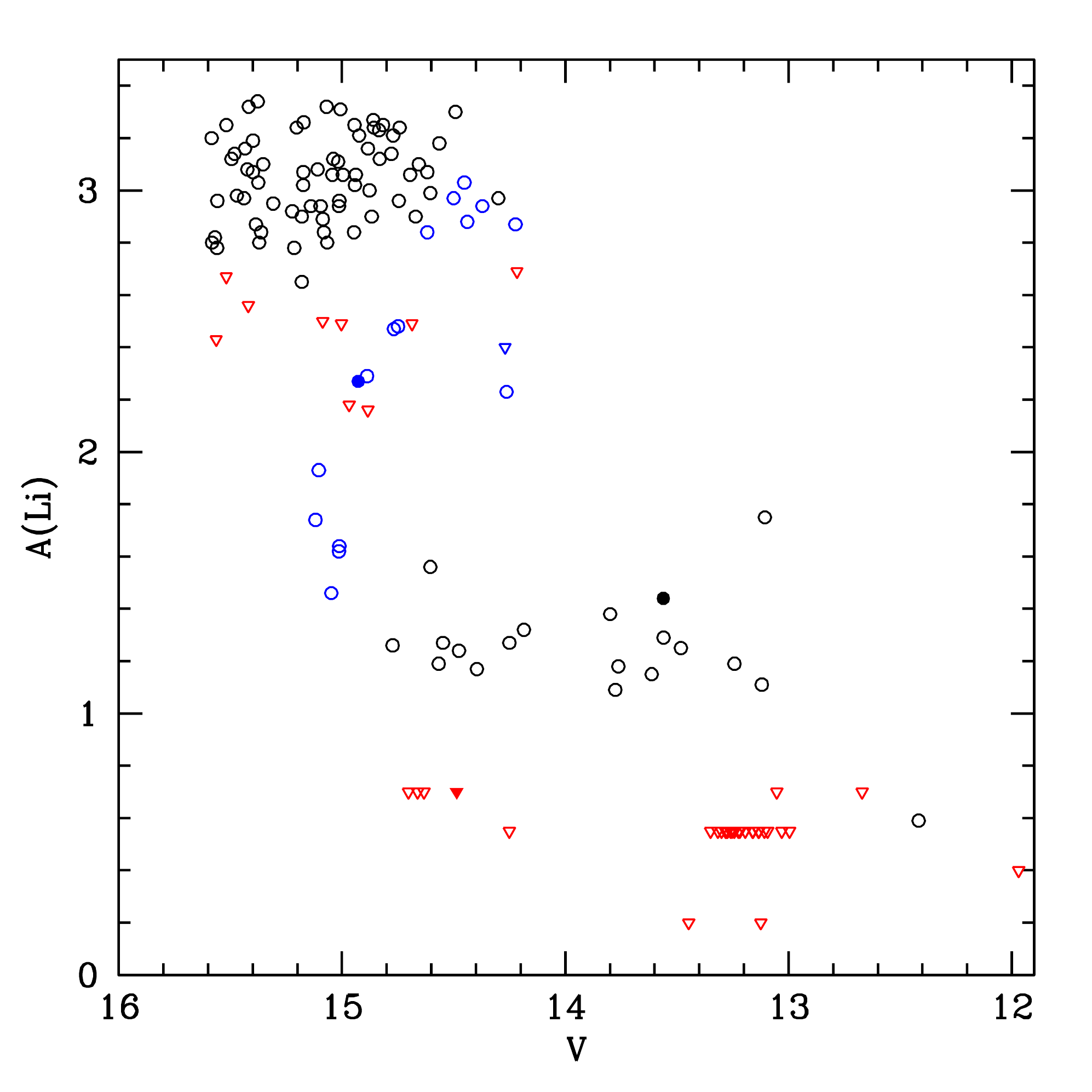}
\caption{Li abundance as a function of $V$. Symbols have the same meaning as in Fig.5.}
\end{figure}

A clearer vision of the evolutionary path is shown in Fig. 7, where A(Li) is plotted as a function of the $B-V$ 
color. For stars at the turnoff, there is again no trend of measured A(Li) with color between $B-V$ = 0.3 and 0.5.
However, once the stars initiate evolution across the subgiant branch, there is a steady decline to a typical 
detection value near A(Li) $\sim$ 1.25, followed by no significant decline from the base of the giant branch 
to the level of the clump. With the exception of only 4128, all stars readily identifiable as post-He-flash 
red giants exhibit only upper limits to A(Li). 

\begin{figure}
\figurenum{7}
\plotone{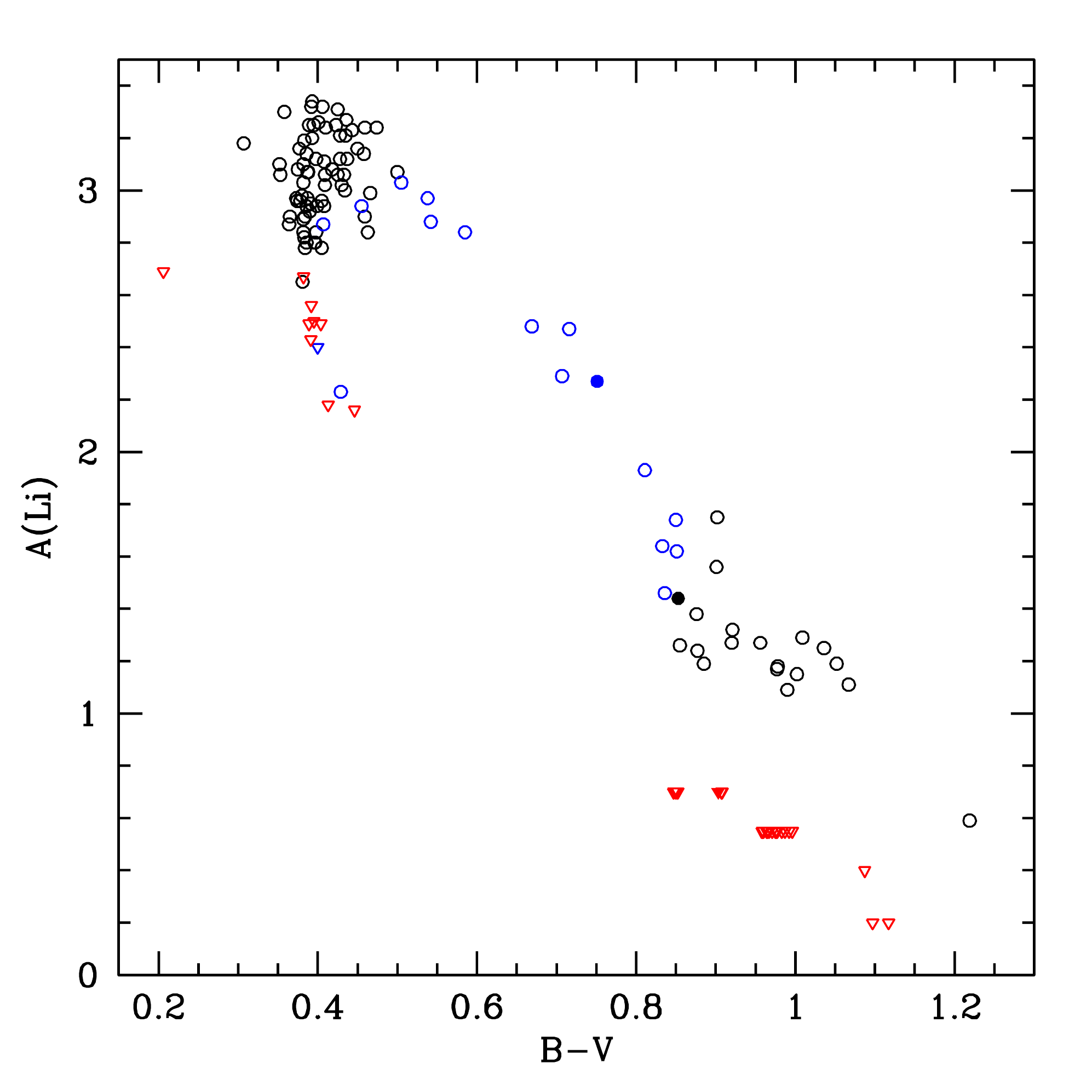}
\caption{Li abundance as a function of $B-V$. Symbols have the same meaning as in Fig.5.}
\end{figure}

What can we learn from these patterns, especially in light of the predictions from standard stellar evolution
models? Beginning with the turnoff, there are 72 stars with $B-V$ $\leq$ 0.50 with detectable Li. The average 
A(Li) for these stars is 3.04 $\pm$ 0.19 (s.d.) dex; if one extreme outlier with A(Li) = 2.23 is excluded, 
the average becomes 3.05 $\pm$ 0.16 (s.d.); the dispersion is more than double the value of 0.07 derived from the
average standard deviation as defined by the precision of the individual measures. The stars at the turnoff of 
NGC 2506 started on the unevolved main sequence hotter than the Li-dip, 
which is located at a fainter $V$ than accessed by the spectroscopic sample. Under standard stellar evolution 
models, the turnoff stars should retain their initial A(Li) until they evolve to $T_{\mathrm{eff}}$ $<$ 5600 K, 
the site of the first signature of the first-dredge-up (FDU) phase, equivalent to $(B-V)$ $\sim$ 0.75 for the 
subgiant stars in NGC 2506, i.e. until one reaches the reddest stars on the subgiant branch (blue symbols) 
at the base of the vertical FRG branch \citep[e.g.,][]{PI97, CH10}. 

\subsection{Initial Conditions: The Primordial Cluster Li Abundance}
While it is generally agreed that the location of the Li-dip is purely temperature-dependent, leading to 
mass ranges for the Li-dip strongly correlated with [Fe/H] \citep{BA95, CH01, AT09, CU12, RA12}, a distinctly 
different question is the evolution of the Li abundance with time and [Fe/H] within the Galaxy. 
Did clusters with lower [Fe/H] form with a lower initial A(Li) or is the time of formation the determining factor? 
This question is independent of the discrepancy between the Li abundance among globular cluster stars and the
primordial estimate from cosmology \citep{CO14} since both lie at A(Li) = 2.7 or less, well below the solar system 
\citep{AS09} and young cluster estimate of $\sim$3.3 \citep{JO97, SD04, BA10, CU17}. The relevance for the
current discussion is that the determination of the physical processes producing a spread of 0.6 dex in A(Li)
among the stars at the turnoff will be heavily weighted by whether the current observed upper bound in A(Li) was the
initial cluster value and all deviations from that value are due to Li depletions or the primordial cluster 
value was lower and the spread is a combination of both enhancements and depletions of Li relative to the mean.

The difficulty in defining the original cluster value arises from the fact that for stars within the Li-dip and
cooler, A(Li) declines over time once stars have reached the main sequence. While the rate of decline will vary
with $T_{\mathrm{eff}}$, with hotter stars on the red side of the Li-dip depleting at a slower rate, unless stars are 
observed relatively soon after attaining the main sequence, the observed abundance supplies only a lower bound.  This
approach has been adopted by investigations attempting to link globular cluster Li observations to the disk clusters and field stars as in \citet{DO14} or from analyses of field star samples of the thin and thick disk, as recently
exemplified by \citet{FU17}. While such discussions invariably conclude that A(Li) has grown between the formation of
the metal-deficient globulars and the current disk, the relative role of age versus metallicity in defining the
growth remains obscure. As already noted, open cluster studies demonstrate that A(Li) depletion for stars cooler than
6500 K is a strong function of temperature and age, even for stars populating the Li plateau redward of the Li-dip \citep{CU12}. The discussion by \citet{FU17}, as an example, derives the trend of A(Li) with [Fe/H] by averaging the 6 most Li-rich stars in each metallicity bin from a sample of $\sim$300 stars ranging in [Fe/H] from -1 to +0.5, sorted into thick and thin disk components based upon [$\alpha$/Fe] as a function of [Fe/H]. The stars with the highest measured A(Li) come from a $T_{\mathrm{eff}}$ range of 6500 K to $\sim$5500 K. The sample includes stars from the unevolved main sequence to the subgiant branch. Without knowing (a) the selection biases within the original sample, (b) the 
sensitivity of the results to the exact choice of the thick/thin disk boundary, a strong function of [Fe/H], and (c) the exact $T_{\mathrm{eff}}$ and age distribution for each star used in defining the upper bound with age, the significance of their derived trend remains questionable. Since no [Fe/H] bin attains an A(Li) upper limit above 2.85, it is clear that the dominant majority of the sample has depleted A(Li) from whatever value they formed with to their current level and attempting to predict the original value which defines the trend with age or [Fe/H] becomes a futile and inherently biased exercise.

From standard stellar evolution theory (SSET), the solution to this constraint would be to observe stars hotter than the
Li-dip since these stars should undergo no depletion because their convective atmospheres are thin to 
non-existent.

The obvious problem with this approach to testing the evolution of A(Li) with age in the Galaxy is that, except 
for star clusters with metallicities well below solar, by an age of 3-4 Gyr the majority of stars hotter than 
the Li-dip have evolved off the main sequence and Li-dip stars are populating the turnoff region, as 
illustrated by M67 and NGC 6253 \citep{CU12} and Ber 32 \citep{RA09}. This clearly mass-dependent transition 
from stars with $normal$ Li abundance on the main sequence to those which are guaranteed to leave the main 
sequence already exceptionally depleted in Li makes comparisons between the giant branches of clusters younger 
than $\sim$2.5 Gyr with those older than this (the exact boundary is, again, metallicity-dependent) 
generally meaningless without first renormalizing the Li scale for the older, lower mass stars.

A cluster which is more metal-poor than NGC 2506 with [Fe/H] $\sim$ -0.5 and somewhat older (3-4 Gyr) is NGC 2243 
\citep{AT05, VA06, JA11, FR13}. Unfortunately, the cluster sits tantalizing close to the boundary where the hot edge 
of the Li-dip is barely contained within the turnoff region. The majority of the stars with published A(Li)
\citep{HI00, FR13} approaching the top of the turnoff have only upper limits to their A(Li), as expected
for Li-dip stars. There are, however, 5 stars extending toward the subgiant branch which have measurable Li, with 
values ranging from A(Li) = 2.92 to 2.39, leading \citet{FR13} to adopt A(Li) = 2.7 $\pm$ 0.2 as the primordial
Li abundance of the cluster. If, instead, the two largest values are adopted as indicative of the original cluster
abundance, A(Li) = 2.9. Preliminary analysis of a much larger sample of turnoff and subgiant spectra obtained with
HYDRA during the same observing cycle as NGC 2506 extends the detection limit for turnoff members of NGC 2243 to
A(Li) above 3.0, with the majority of stars covering the range between 2.8 and 2.3. Statistical issues aside, since NGC 2243 
is both older and more metal-poor than NGC 2506, the apparent reduced Li boundary for the hot side of the turnoff doesn't 
supply any insight into which parameter, time of formation or metallicity, has greater influence on the 
primordial A(Li) value.

A similar problem applies to the more metal-rich ([Fe/H] = -0.02 \citep{LB15}) cluster, NGC 6819, with an age of 2.25 Gyr 
\citep[][in prep.]{DE18}. Although it is much closer in age to NGC 2506 than NGC 2243, the higher metallicity places the Li-dip at 
a higher mass than in NGC 2506, and thus the stars brighter than the Li-dip are, on average, more evolved than the stars
brighter than the Li-dip in NGC 2506. The apparent result is that the stars brighter than the Li-dip at the turnoff in NGC 6819 exhibit 
a much wider range of A(Li) than found among the turnoff stars in NGC 2506. As we will discuss below, the significant 
spread, from an upper limit for single stars of A(Li) = 3.2 to detections below A(Li) = 2.0, and non-detections for 
a majority of the stars, is a clear indication that some parameter other than $T_{\mathrm{eff}}$ must serve as a catalyst 
for Li-depletion upon leaving the main sequence.

Perhaps the best analog to NGC 2506 is IC 4651, a cluster consistently observed through both photometry and spectroscopy
to be above solar metallicity with [Fe/H] typically between +0.10 and +0.15 \citep{AT00, ME02, PA04, CA04, SA09}. 
Its younger age than NGC 2506 compensates in part for the shift to higher mass for the location of the Li-dip, placing 
the stars near the turnoff in a position relative to the Li-dip similar to that of NGC 2506. The turnoff region of the 
cluster is tight and well-defined, as in NGC 2506, and it is one of the few clusters in the 1-2 Gyr age range to have 
a few stars populate the subgiant branch just beyond the turnoff. While the sample of stars observed for Li  (and Be) 
to date is modest \citep{BA91, PA04, SM10}, the composite Li data for the cluster at higher mass than the Li-dip show 
a maximum A(Li) between 3.3 and 3.4 before dropping precipitously across the subgiant branch (see Fig. 9 of \citet{AT09}). 
Unfortunately, the full range of stars at the turnoff brighter than the Li-dip extends down to A(Li) $\sim$ 2.5, a spread confirmed 
using the composite sample from NGC 752, NGC 3680, and IC 4651 \citep{AT09}. 

Looking to clusters younger and more metal-rich than NGC 2506, the obvious choices are the virtually identical
clusters, the Hyades and Praesepe, as recently investigated by \citet{CU17}. With [Fe/H] = +0.15 and ages less than
1 Gyr, the combined sample should provide a reasonable test of any correlation between A(Li) and [Fe/H] at a
given $T_{\mathrm{eff}}$ or main sequence mass. As illustrated in Fig. 13 of \citet{CU17}, from the small composite 
sample using only single-star cluster members, A(Li) does rise to a plateau value of A(Li) $\sim$ 3.3 just hotter 
than the Li-dip boundary. Among the hotter A stars, however, the Li abundance declines, reaching A(Li) $\sim$ 2.5.

Perhaps the best evidence for a Li-Fe correlation comes from an analysis of Hyades-age clusters ranging in [Fe/H]
from +0.15 to -0.23 by \citet{CU08}. In addition to the already mentioned Hyades and Praesepe sample at [Fe/H] = +0.15,
A(Li) as a function of $T_{\mathrm{eff}}$ from $\sim$7000 K to 4500 K was measured for a large sample of stars in
NGC 2539 ([Fe/H] = 0.00), IC 4756 ([Fe/H] = -0.10), NGC 6633 ([Fe/H] = -0.10), and M36 [Fe/H] = -0.23. Each cluster exhibited a well-defined trend between A(Li) and $T_{\mathrm{eff}}$, with virtually identical patterns for paired clusters of the same metallicity. However, when the relations at the three metallicities were superposed, they had distinctly different 
zero-points and slopes. For $T_{\mathrm{eff}}$ $>$ 5700 K, A(Li) declined with decreasing [Fe/H] while for the cooler
sample, the pattern reversed. The solution to the contradiction with the theoretical prediction from main sequence 
models with mixing due to convection is twofold: first, adopt a primordial cluster A(Li) strongly correlated with [Fe/H], i.e. the Hyades and Praesepe formed with A(Li) higher than that found in IC 4756 and NGC 6633 by $\sim$0.3 dex. Based
upon the relation derived from cluster data \citep{CU08}, a straightforward, unweighted linear fit between 
A(Li) and [Fe/H] over the [Fe/H] range from -0.2 to +0.1 gives:

\begin{eqnarray}
\nonumber
A(Li) = 3.315 \pm 0.003 +  0.959 \pm 0.034[{\rm Fe/H}]
\end{eqnarray}

\noindent One would expect a cluster with [Fe/H] = -0.27 like NGC 2506 to have a primordial value of A(Li) = 3.06, consistent with the mean value observed among the turnoff stars more massive than the Li-dip. Second, for stars cooler than
the Li-dip, the rate of Li destruction at a given $T_{\mathrm{eff}}$ is metallicity dependent, with more metal-rich clusters exhibiting higher rates of depletion.

Collectively, the evidence points (a) with high probability to the fact that the scatter in A(Li) among the stars 
at the cluster turnoff is real and a product of evolution on the main sequence and (b) to the possibility that the 
lower mean cluster Li abundance in NGC 2506 (A(Li) = 3.0) compared with clusters with more Hyades-like 
metallicity (A(Li) = 3.4) is real and is tied to the intrinsically lower [Fe/H] of the former. 
Just how much NGC 2506 and NGC 2243 lie below the more metal-rich objects ultimately depends on how one weights 
the wide distribution of A(Li) found among the sample of stars more massive than the Li-dip when deriving the 
mean A(Li). Particularly for clusters older than the Hyades, there is growing evidence that A(Li) above the 
Li-dip shows a more extensive range, especially toward lower A(Li), as a cluster ages \citep[][in prep.]{DE18}. Equally concerning
is the observational fact that the range in A(Li) for these samples, irrespective of the cluster [Fe/H], exhibits 
a similar upper bound approaching A(Li) = 3.35. We will return to this point after discussing the trends in 
A(Li) among stars beyond the cluster turnoff.

\subsection{Evolution on the Subgiant Branch and Beyond}

The second empirical insight gained from the A(Li) measures comes from the color evolution illustrated in Fig. 7. 
There is a clear range of A(Li) between 2.75 and 3.35 for the stars at the turnoff with $B-V$ bluer than 0.5 but 
no statistical evidence for a variation with temperature/color or $V$. For stars with $B-V$ redder than
0.5, there is a clear decline in A(Li) as stars evolve across the subgiant branch, reaching a typical
abundance of A(Li) $\sim$ 1.25 at the base of the red giant branch, near the expected location for the end of the
FDU phase. The pattern of a distinct transition in A(Li) as a function of $T_{\mathrm{eff}}$ is reminiscent of the 
Li-dip on the blue/hotter side among less evolved main sequence stars. From the precision composite data for young 
clusters \citep{CU17}, intermediate-age clusters \citep{AT09, CU12}, and field stars \citep{RA12}, stars on the 
unevolved main sequence which were hotter than $T_{\mathrm{eff}}$ $\sim$ 6700 K exhibit well-defined values of 
A(Li) between 3.2 and 3.4 or, at minimum, a range of A(Li) extending as high as these limiting values. For 
stars at $T_{\mathrm{eff}}$ $\sim$ 6600 K and as old as or older than the Hyades, one can only find stars 
with minimal or undetectable A(Li), defining the approximate center of the Li-dip. This $T_{\mathrm{eff}}$ 
boundary is readily seen in intermediate-age and older clusters where the unevolved main sequence temperature 
pattern translates into a break at a specific $V$ magnitude due to the evolution of the key mass range into 
an approximately vertical turnoff (see, e.g. Fig. 10 of \citet{AT09}). 

If the physical mechanism defining the high mass boundary of the Li-dip is predominantly defined by the $T_{\mathrm{eff}}$ 
of the star, stars of higher mass which form with temperatures hotter than this limit might be expected to initiate 
Li-depletion upon crossing this boundary during post-main-sequence evolution toward and across the subgiant branch.
For stars at the turnoff of NGC 2506, with $E(B-V)$ = 0.06, the color boundary for $T_{\mathrm{eff}}$ = 6700 K
should be $B-V$ = 0.43. While the sample is small, the majority of stars between $B-V$ = 0.43 and 0.5 do not show a
dramatic drop in A(Li); the boundary of $B-V$ = 0.5 is equivalent to $T_{\mathrm{eff}}$ = 6450 K, placing it already
beyond the center of the Li-dip defined by lower mass stars. In short, dwarfs exhibit a distinct Li-dip while 
higher-mass subgiants in the same $T_{\mathrm{eff}}$ range do not. Since the Li-dip among dwarfs requires a 
few hundred million years to develop, the lack of a distinct boundary could arise from the more rapid evolutionary 
timescale for stars crossing the subgiant branch. This immediate lack of coupling by stars at the turnoff and the Li-dip 
temperature boundary is even more evident in the slightly older cluster, NGC 6819, where almost the entire turnoff 
region brighter than the Li-dip lies redward of the Li-dip $T_{\mathrm{eff}}$ boundary but these higher
mass stars retain a range of A(Li) extending to 3.2 \citep[][in prep.]{DE18}. Therefore, while $T_{\mathrm{eff}}$ can 
play a valuable role in marking the boundary for the initiation of extra mixing and/or Li dilution not predicted 
by SSET for stars on or leaving the main sequence, it supplies no particular insight into the physical mechanism 
driving the process of depletion. 

\begin{figure}
\figurenum{8}
\plotone{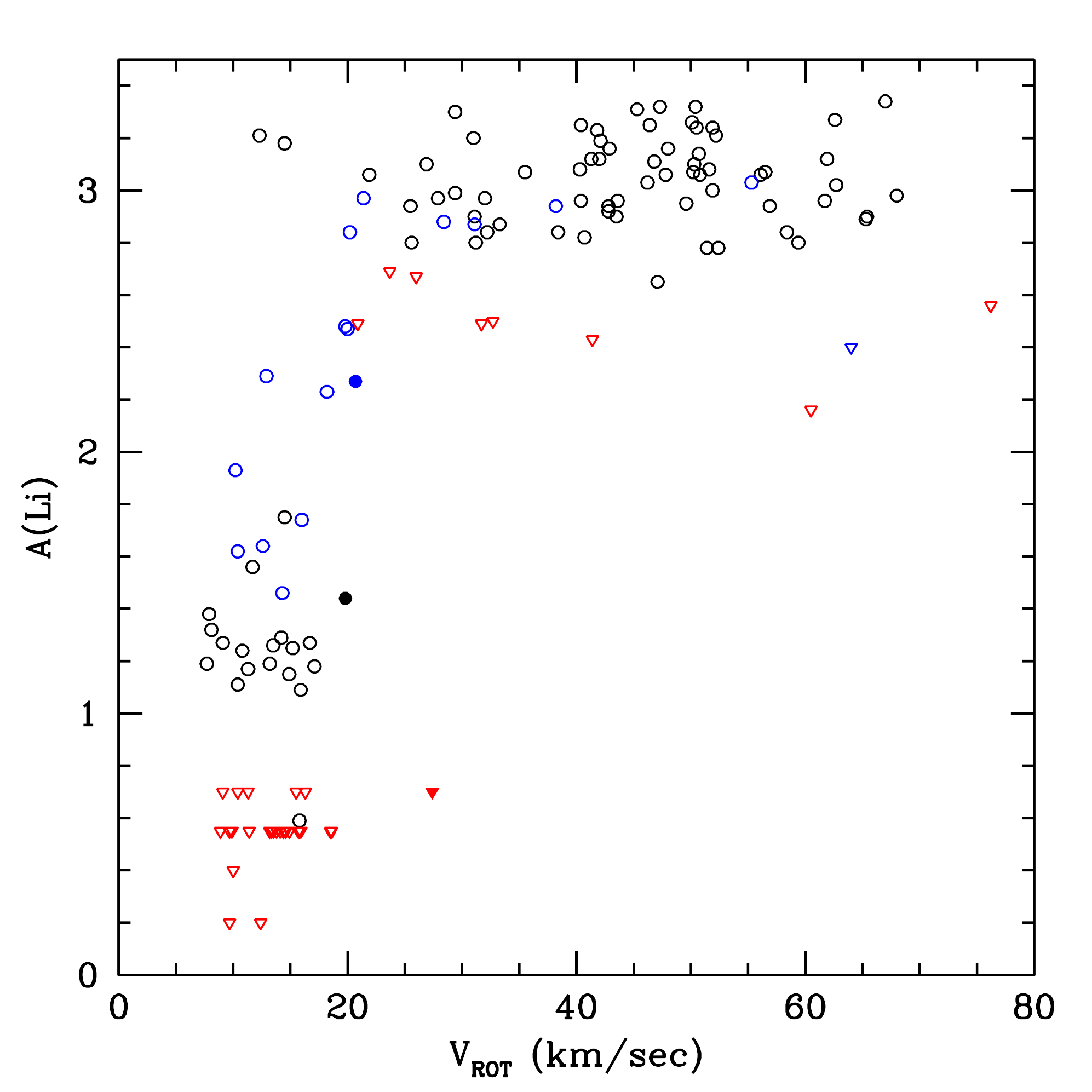}
\caption{Li abundance as a function of rotational speed. Symbols have the same meaning as in Fig.5.}
\end{figure}

An alternative possibility emerges from the other significant stellar property which changes decisively for stars 
entering and crossing the subgiant branch, $V_{rot}$. Fig. 8 shows the trend of A(Li) as a function 
of $V_{rot}$ for the stars used in Figures 5, 6, and 7. Symbols have the same meaning as in the previous figures. 
First, there is no apparent correlation between A(Li) and rotational speed. As discussed earlier, the majority of 
stars for which reliable A(Li) estimation was deemed implausible were rapid rotators ($V_{rot}$ $>$ 30 km s$^{-1}$). A
significant concern for such rotators is that the increased blending of the neighboring lines makes 
it challenging to judge the continuum level and account for line contamination. Unlike dwarfs in
the Hyades and Praesepe \citep{CU17}, however, a metallicity less than half that of the younger clusters makes
this less of an issue and clearly introduces no trend in the data for stars where A(Li) is detectable.

Second, keeping in mind that the $sin$ $i$ factor can move any star toward a lower than true rotation speed, i.e. 
to the left in Fig. 8, the sharp transition near $V_{rot}$ = 25 km s$^{-1}$ is striking. Stars with A(Li) 
covering the range from 2.75 to 3.35 can have virtually any rotation speed from 12 to 65 km s$^{-1}$; by contrast, 
with only one exception, every star with A(Li) below 2.4 has $V_{rot}$ below 25 km s$^{-1}$. 

The significance of the distribution of A(Li) with $V_{rot}$ among turnoff stars and giants becomes apparent when NGC 2506
is placed in the context of clusters ranging in age from 1.4 Gyr (NGC 7789) to NGC 3680 (1.7 Gyr) to NGC 6819 (2.25 Gyr).
As detailed in \citet[][in prep.]{DE18}, for stars brighter than the Li-dip at the cluster turnoff, the range in $V_{rot}$ extends
above 100 km s$^{-1}$ for the youngest cluster, declines to $\sim$60 km s$^{-1}$ for the intermediate-age cluster, and drops
to $\sim$25 km s$^{-1}$ for the oldest. NGC 2506 therefore most resembles NGC 3680, taking into account the lower mass
of the Li-dip at lower [Fe/H]. Equally relevant, as the cluster age rises, the fraction of stars at the turnoff and
more massive than the Li-dip stars with depleted or undetectable Li rises from less than 15\% to more than 70\%. This 
trend also translates into a dramatic evolution of the Li among the giants. While more than half of the evolved stars 
in NGC 2506 have detectable Li, only 7 of the 51 stars classed as subgiants or giants in NGC 6819 have detectable Li \citep[][in prep.]{DE18}. 

Before continuing with the distribution of A(Li) among the giants on the FRG branch, we can ask
how well the empirical observations of Li evolution from the main sequence through the subgiant branch agree with  
the predictions of standard stellar evolution theory, in the absence of two critically important processes, 
rotation-induced mixing and thermohaline mixing. The general answer is that, on many details, they don't agree. As
first emphasized by \citet{PA04} using the models of \citet{CH99, PA03} applied to the sparse data for IC 4651,
models without rotation are incapable of explaining any decline in A(Li) among intermediate mass 
stars (M = 1.8 M$_{\sun}$) on the main sequence and beyond until the FDU, well across the Hertzsprung gap, 
in contradiction with the cluster data. By contrast, stellar models with the same mass but rotating above 
100 km s$^{-1}$ show significant depletions in atmospheric Li initiated at temperatures much hotter than in the 
non-rotating stars and comparable to the $T_{\mathrm{eff}}$ of stars just leaving the main sequence.

The analysis of IC 4651 was revised using the newer models of \citet{CH10} to include rotation-induced
mixing, internal gravity waves, atomic diffusion, and thermohaline mixing and expanded to include the evolutionary
pattern for Be, once again confirming the need for additional mixing processes beyond simple convection to explain
the abundance patterns observed in field stars and clusters \citep{SM10}. With the exceptionally well-defined
trend of A(Li) across the subgiant branch of NGC 2506, we can more effectively test the model predictions using the discussion 
of \citet{CH10}. The closest model analog to the stars at the turnoff and along the giant branch of NGC 2506 
which have all three variants: standard evolution, standard with thermohaline mixing, and standard with thermohaline 
mixing and rotation are the 1.5 M$_{\sun}$ models with rotation speeds near 0 or 110 km s$^{-1}$. Note that the model 
metallicity is solar and the rotation speed is larger than observed for the stars in NGC 2506, but our 
primary interest is in the qualitative pattern.

If we use standard evolution with or without thermohaline mixing, non-rotating stars leaving the main sequence retain 
their primordial A(Li) until they are almost 1600 K cooler than the turnoff $T_{\mathrm{eff}}$, or $(B-V)$ = 0.75 for
NGC 2506. At the hottest point of the turnoff, the surface convection zone (SCZ) occupies a tiny fraction by mass of 
the outermost layers. As the model evolves to cooler $T_{\mathrm{eff}}$ along the subgiant branch, the SCZ deepens 
substantially and continuously. In standard theory (e.g. \citet{DE90}), the Li abundance as a function of depth has 
not been altered substantially while on the the main sequence until a depth is reached where Li is broken apart by 
energetic protons, the Li preservation boundary.  Below that boundary, Li declines steeply with depth.  Therefore, 
the surface Li abundance stays constant until the SCZ deepens sufficiently to reach the Li preservation boundary.  
As noted earlier, this happens at $T_{\mathrm{eff}}$ $\sim$ 5600 K.  Further evolution begins to reveal the impact 
of the first dredge-up, as the SCZ deepens into regions where Li was destroyed during the MS and pre-MS phases and 
are thus now devoid of Li.  Even as the SCZ deepens, the model expands and these layers are now much cooler than 
they were during the MS: no further Li destruction occurs at the base of the SCZ.  Instead, convection mixes the 
outermost regions that still contain Li with those regions that are devoid of Li, resulting in a decrease of the 
surface Li abundance, i.e., subgiant dilution.

This decrease is a steep function of decreasing $T_{\mathrm{eff}}$ because by the time the SCZ reaches the base of the FRG 
branch, it includes more than half of the stellar mass. By contrast, the Li preservation region occupies only a 
small fraction of the stellar mass, so the total dilution of surface Li is roughly 1.8 dex \citep{DE90, CH10}.  
The SCZ reaches a maximum depth near the base of the FRG branch at $\sim$4800 K, after which the base of the SCZ 
recedes slowly toward the surface, as the model evolves up the FRG branch.  Standard theory predicts no further 
depletion of surface Li up the giant branch, thereby establishing the {\it subgiant diluted Li plateau}, namely 
a constant Li abundance up the FRG. If the turnoff A(Li) $\sim$ 3.3, then the diluted plateau is predicted to have 
A(Li) $\sim$ 1.5. If A(Li) of the star leaving the turnoff is lower, either due to a lower primordial Li value 
associated with a lower metallicity cluster and/or due to non-standard Li depletion on the main sequence, then the 
diluted plateau A(Li) could easily approach A(Li) = 1.2. This would be consistent with the plateau of Li abundances 
seen in Figure 6 and 7. But what about those stars with A(Li) lying substantially below the diluted Li plateau?

Subgiant evolution can show the effects of non-standard surface Li depletion mechanisms that might have acted 
during the MS and beyond, and that might have reduced the amount of Li in the Li preservation region. In the case of 
the models of \citet{CH10}, inclusion of rotation and rotationally-induced mixing produces an immediate and 
continuous decline in A(Li) as soon as the star begins evolving away from the main sequence toward the giant branch. 
By the end of the first dredge-up phase the atmospheric A(Li) is reduced to less than 0.5. It is expected that adoption of a
lower initial rotation rate would generate the same trend, but approach a higher limit for A(Li) beyond the first dredge-up.
Exactly how the final value for A(Li) correlates with increased $V_{rot}$ requires extensive modelling beyond those
currently available.

Moving beyond the base of the FRG branch, the deficiency in Li among the clump stars is 
predictable since all these stars have supposedly undergone He-flash, generating mixing which could reduce 
and/or eliminate any remaining signs of the element in their atmospheres as they arrive at the red giant tip.
While burning He in a stable configuration on the clump, the A(Li) should remain unchanged until the star begins
its ascent up the asymptotic giant branch and approaches the second dredge-up phase.
 
A more uncertain interpretation applies to the concentration of Li-deficient stars at the base of the giant branch.
The key structural feature which kicks in near this phase of evolution is the FDU. As already noted, if the surface 
convection zone reaches deep enough, it can mix atmospheric Li with Li-depleted layers well below the Li-preservation
zone, potentially causing a sudden  and significant decrease in the observed A(Li), as seen in Fig. 5 near $V$ = 14.8. 
If this is the case, why then do the majority of stars on the FRG branch between $V$ = 14.6 and 13 in Fig. 5 
have measurable and {\it constant} A(Li)?

One potential solution is that the difference between measurable and detectable Li among the spectra is small enough that
only a slight change in the strength of the line will shift a star from one category to the other, i.e. the bimodal
Li distribution among these giants is an artifact of the spectra. To illustrate why this fails, we show in Fig. 9 a comparison
of two stars at the base of the giant branch with virtually identical colors and magnitudes, star 1301 with A(Li) near 1.56 and 
star 7008 with only an upper limit near A(Li) = 0.6. It is clear that there is a distinct difference in the strength of the
Li line for these two stars, consistent with their classification, even though all other lines have the same strength within
the spectroscopic uncertainties.

\begin{figure}

\figurenum{9}

\plotone{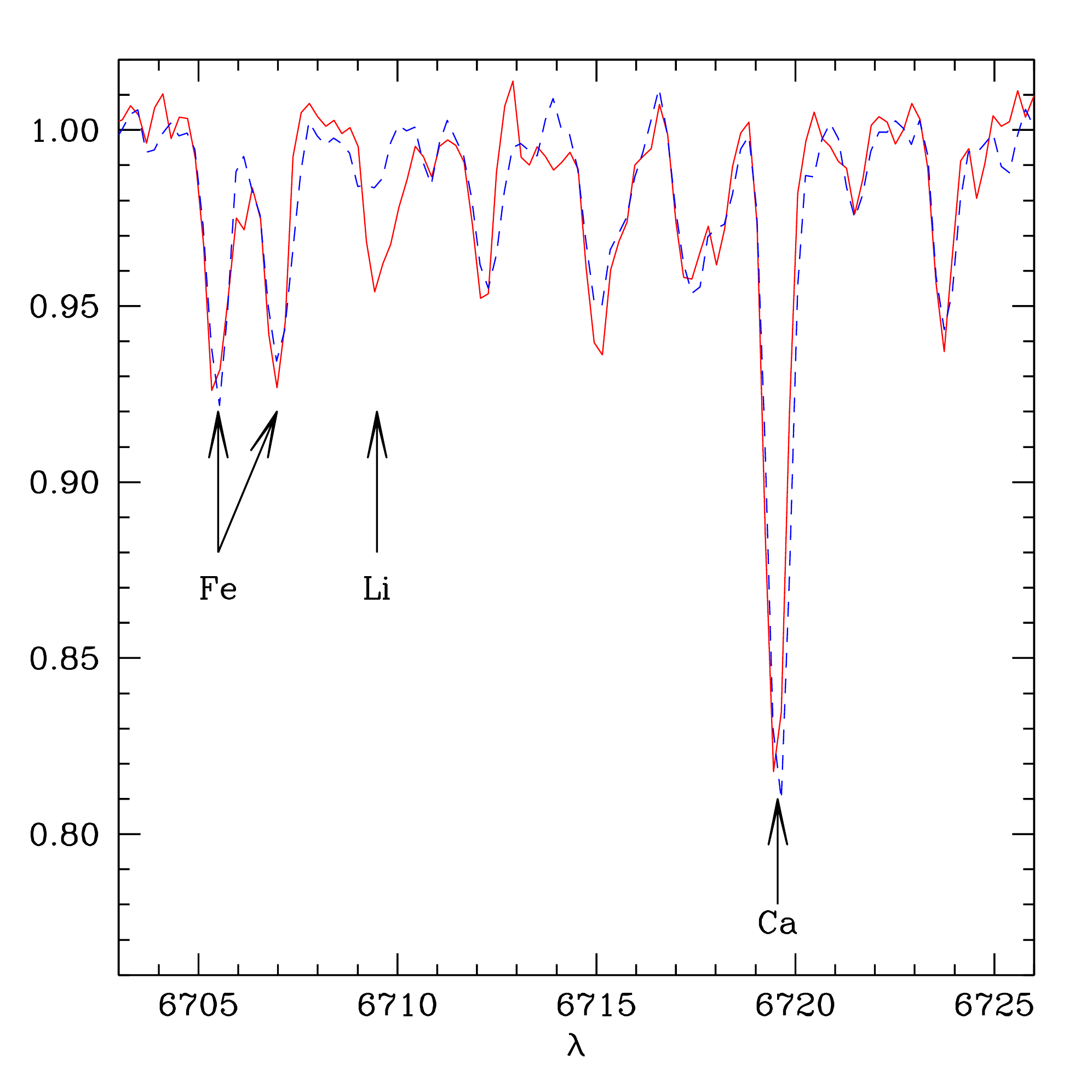}
\caption{Closeup of Lithium line region in two stars at the base of the 
giant branch. The red spectrum traces star 1301, the blue spectrum star 7008. A number of lines in the Li region are identified.}

\end{figure}

Assuming that the majority of these stars are members and do not represent a low-luminosity extension of the red giant clump, two plausible options exist: (a) these stars aren't single stars, but instead are evolved binaries/blue stragglers 
approaching the base of the giant branch at a higher luminosity, or (b) the stars with more extreme Li-depletion than the
Li-plateau stars are those which initially had the highest rotation speeds and have been systematically more
affected by the mixing processes controlled by rotation and stellar spindown.  We reiterate that one of the stars among 
this group has a spectroscopic [Fe/H] near solar indicating that, despite the radial velocity agreement with the 
cluster value, it may be a field interloper.

From Fig. 7, among the stars with measurable Li, there is little evidence for a decline in A(Li) as one moves up the
giant branch to the level of the clump. Instead, from 16 FRGs (open circles in Figs. 5, 6, and 7), the stars have 
A(Li) = 1.25 with a scatter of $\pm$ 0.11 dex. If one star near the base of the FRG is excluded, the average becomes 
1.22 $\pm$ 0.08, only slightly higher than expected from the precision of the measures, but significantly lower than found among 
the stars at the turnoff. For SSET and for models with rotational mixing on the main sequence, A(Li) for a star is predicted 
to remain constant at the value it has attained after completion of the FDU. 

The next key change predicted in the surface A(Li) occurs once the stars pass beyond the red giant bump. The red giant bump 
represents an evolutionary slowdown and reversal as the hydrogen-burning shell passes across the chemical composition discontinuity 
created by the high-water mark of the convection zone as the stars evolved toward and up the giant branch. The change in 
molecular weight and H concentration allows the giant to support itself at a lower luminosity and then evolve back up the 
giant branch at a slower rate, causing a density excess in the distribution of giants. Standard models predict that stars 
evolving beyond the bump will experience no change in A(Li) until they reach the tip of the giant branch. By contrast, if 
one includes thermohaline mixing, stars brighter than the bump will decrease their lithium by 0.4 dex. The location of 
the red giant bump in NGC 2506, predicted by
VR isochrones with age between 1.8 and 1.9 Gyr and [Fe/H] = $-0.29$, is 
between $V$ = 12.95 and 12.7, above the observed level of the clump. Clearly, there is no excess concentration of stars 
in this magnitude range in Fig. 5. However, the \citet{CH10} models for the 1.5 M$_{\sun}$ with rotation 
($V_{rot}$ = 110 km s$^{-1}$) demonstrate that the presence of rotation will shift the location of the red giant bump 
and the start of thermohaline mixing to a lower luminosity on the giant branch, typically 40$\%$ fainter for the bump and 
a factor of almost 8 for the mixing trigger, placing the mixing phase only 3-4 times brighter than the red giant bump. 
While the size of the change will depend upon the specific size of the rotation, inclusion of rotation should move the 
bump fainter than the level of the red giant clump. Equally important, additional depletion of Li caused by thermohaline 
mixing should begin to show among stars on the FRG branch no more than a magnitude brighter than the clump. Thus, the redefinition of the FRG
branch passing through star 2402 with detectable Li just below A(Li) = 0.6 is consistent with a significant decline from
A(Li) = 1.25 caused by mixing in a star beyond the red giant bump. The other stars brighter than 2402 at comparable colors 
would then be either binaries undergoing peculiar evolution or stars ascending the asymptotic giant branch.

\section{Summary and Conclusions}

HYDRA spectroscopy of 287 stars in the field of NGC 2506 has been used in conjunction with published proper-motion 
membership and position in the CMD to identify highly probable, single-star members of the cluster for further analysis.
The survey included 24 stars within the previously studied area of the cluster and 135 stars outside the cluster area
for which the only available information was the location in the CMD. Of these 159, 94 proved to be probable radial-velocity 
members, adding a critical number of stars to evolutionary phases of the CMD which are normally poorly, if at all, populated.
Abundance analysis for metallicity determination, dominated by the richer and stronger line profiles of the cooler stars,
confirms that NGC 2506 is moderately metal-poor, with [Fe/H] near -0.3.  

Returning to the primary focus of the investigation, the evolution of Li within the cluster as a function of mass, a number
of conclusions stand out: 

There is an intrinsic spread in A(Li) among the stars at the vertical turnoff, though the range is independent of 
stellar luminosity or color. Since all the observed turnoff stars lie brighter than the Li-dip and, by SSET, 
have retained their initial Li abundance unchanged by evolution, the mean A(Li) for the cluster is calculated to be 
3.05 $\pm$ 0.16, consistent with a lower initial abundance than the sun \citep{AS09}, as predicted for cluster which is 
a factor of two below solar in [Fe/H]. It is noted, however, that the full range of the sample in A(Li) extends from 
2.75 to 3.35, comparable to that found among stars blueward of the Li-dip in clusters of solar metallicity or higher.   

Upon exiting the turnoff region and evolving across the subgiant branch, the stars undergo a well-delineated
decline in atmospheric Li, declining from a mean of A(Li) = 3.05 to a plateau value of 1.25 by the base of the FRG branch. 
While it has been obvious for decades that the typical red giant contains less Li than expected for a star
of the same mass on the main sequence, it has been a challenge to identify precisely when and where the depletion occurs.
Stars fainter than the clump in clusters of intermediate age exhibit depleted Li, but the degree of depletion can vary by
a factor of ten and it can be difficult to assess if the star is truly a first-ascent or red clump giant. 

We reiterate that the evolutionary phase under discussion is not the vertical turnoff region. 
Intermediate-age clusters like NGC 3680 and NGC 752 do show a spread in A(Li) ranging from 3.3 to 2.6 at 
the top of the vertical turnoff \citep{AT09}, virtually identical to the range found in NGC 2506. 
IC 4651 exhibits the same pattern, but is the one cluster in the 1-2 Gyr age category to include multiple 
stars populating the subgiant branch between the vertical turnoff and the red giant branch. 
\citet{PA04} have studied a limited sample of stars in IC 4651 and find four at the top of the turnoff and 
one at a temperature intermediate between the turnoff stars and the red clump which have A(Li) between 
2.4 and 1.6. Note that the one star with a $T_{\mathrm{eff}}$ placing it in the subgiant
region has A(Li) = 2.1. For the first time, the mapping of a subgiant branch using stars outside 
the Li-dip demonstrates that these stars continuously deplete Li as they evolve to the base of the FRG branch, 
irrespective of whatever mechanism, if any, produces the dispersion among stars at the top of the turnoff.

A contributing factor leading to the onset Li-depletion on the subgiant branch appears to be the spindown of 
the stars as they evolve to cooler temperatures at almost constant luminosity. Stars at the vertical turnoff 
within the A(Li) range discussed earlier can have any rotation speed between $\sim$10 km s$^{-1}$and 70 
km s$^{-1}$, with the caveat that the true rotation speed could be higher and that the spectra have 
resolution which limits the $V_{rot}$ to a minimum approaching 10 km s$^{-1}$. There is no
correlation between luminosity and/or temperature and $V_{rot}$. By contrast, as stars evolve 
across the subgiant branch and the typical $V_{rot}$ declines from a minimum of 25 km s$^{-1}$ 
to an average near 13 km s$^{-1}$, A(Li) drops steadily to 1.25. This pattern fits perfectly within the 
trend defined by NGC 7789, NGC 3680, and NGC 6819, where the fraction of turnoff and giant stars with 
detectable Li is strongly correlated with the spread in rotation speed among stars hotter than
the Li-dip; the larger the range in speed at a given age, the more likely the turnoff stars will have a modest range in A(Li)
with a mean at or above 3.0, and the more likely that the giants will exhibit detectable Li near 1 \citep[][in prep.]{DE18}.

It should be noted that the $V_{rot}$ distribution for stars brighter than the Li-dip at the turnoff and on the giant branch of IC 4651 \citep{ME02}
is virtually identical to that in NGC 3680. For stars at the turnoff, the range is from $\sim 10$ km s$^{-1}$ to
more than 60 km s$^{-1}$ while the giants exhibit a small scatter near 1 km s$^{-1}$ \citep{ME02}. The uncertainty in
the latter velocity measures is analogous to that for \citet{CA14} rather than our HYDRA data. Unfortunately, when binaries are 
eliminated, the current number of stars with reliable Li determinations brighter than the Li-dip is too small to shed any
statistical light on the question. A comprehensive survey of this cluster with a significant expansion
of the spectroscopic sample could prove invaluable.

The Li pattern exhibited at the turnoff is consistent with the predictions of stellar models of low/intermediate mass
which include rotation-induced mixing on the main sequence and beyond \citep{CH10}. The growing dispersion in Li for stars on
the blue side of the Li-dip as clusters age from 1 to 2 Gyrs then becomes a reflection of the initial spread in $V_{rot}$ 
acting over time to reduce the absolute Li at Hyades age from a uniform value at or above A(Li) = 3.0. This dispersion
is then coupled with the significant spindown of all the stars evolving along the subgiant branch, reducing the Li
abundance to a typical value of 1.2 by the base of the FRG and the end of the first dredge-up. If the decline in A(Li)
at the start of the subgiant branch is not tied to the evolution of the rotation rate but simply defines the start of the
first dredge-up, this phase is triggered at a much higher $T_{\mathrm{eff}}$ (6450 K) than predicted by the 
non-rotating models (5600 K).

A fundamental challenge to this qualitative explanation for the spread in A(Li) among the turnoff stars is presented by
the absence of a comparable spread among the evolved stars of the FRG branch. The dispersion in A(Li) for stars at the
turnoff is twice that among the giants; if stars leave the main sequence with a factor of 4 range in the Li abundance 
due to the range in $V_{rot}$, why doesn't that spread persist among stars on the FRG branch? An alternative solution 
to the dispersion question is presented by diffusion among the metal-deficient stars at the cluster turnoff. 
For very thin SCZs among stars in the $T_{\mathrm{eff}}$ range 6800 K to 7100 K, Li can undergo radiative acceleration 
into the SCZ, creating potentially significant surface overabundances, if the radiative acceleration region is much 
larger by mass fraction than the SCZ \citep{RM93}. If the initial A(Li) for the metal-poor NGC 2506 was $\sim$3.0, 
this mechanism could create surface A(Li) as high as 3.3 or higher, without destroying Li inside the star.  
Inside the star, below the radiative acceleration boundary, Li sinks, with a small amount of it potentially 
reaching depths where some, but not much, might be destroyed. For stars cooler than about 6700 K, the radiative 
acceleration boundary is within the SCZ, so only downward diffusion at the base of the SCZ can occur. The net 
result is that as stars leave the main sequence and the SCZ grows, mixing of the inhomogeneous layers returns A(Li) to the initial 
value of $\sim$3.0 with only a modest dispersion. Beyond that point, whatever
mechanisms exist to deplete Li on the subgiant branch and beyond act on a significantly more homogeneous initial abundance.
Clearly, deciding on the relative viability of either process is beyond the scope of the current observations. 

Moving to the FRG branch, it is tempting to associate the handful of Li-deficient giants at the base of the FRG with 
the FDU as the SCZ accesses deeper and hotter layers below the atmosphere. If the first dredge-up is being outlined 
by the stars with upper limits to Li, it would be beneficial to look for other signatures of mixed processed nuclear 
material, e.g. $^{12}C/^{13}C$ ratios, in the atmospheres of these and the more evolved stars \citep[e.g.,][]{CH10, KA14}. 
To date, $^{12}C/^{13}C$ ratios have been measured for 3 giants by \citet{MI11} and 3 single-star (3265 excluded) giants 
by \citet{CA16}. Five of the 6 stars are located in the clump and one sits 1.5 mag above the clump, classified as a red-giant-tip star 
by \citet{MI11}, though this becomes debatable in light of the CMD position of 2402. Ignoring for now the one
star with a lower limit to the $^{12}C/^{13}C$ ratio of 10, the remaining 5 stars have an average ratio of 10 $\pm$ 2, 
indicating that the stars all have the same $^{12}C/^{13}C$ ratio, within the uncertainties for the individual measures. 
This low value is also consistent with rotating models which include thermohaline mixing, whether the star is at the red giant
tip or completing the second dredge-up phase \citep{CH10}. The models do predict a significantly higher value above 20 for stars
completing the first dredge-up. So, while the $^{12}C/^{13}C$ ratio can't distinguish between a first or second-ascent red giant,
it should provide insight into the status of the Li-depleted giants at the base of the FRG.

For stars with detectable Li on the FRG, within the scatter, there appears to be no evidence for a systematic depletion of Li
up to the level of the red clump. The one star (2402) above the clump with detectable Li, if it defines the path of the 
FRG above the clump, exhibits depletion relative to the other giants which is consistent with a post-bump giant undergoing 
thermohaline mixing. The stars within the clump, with one anomalously blue exception, exhibit only upper limits below the level of star 2402.
For comparison, with four measured clump stars at higher resolution, \citet{CA16} determine a detectable LTE Li abundance 
(the appropriate comparison for our data) for two stars and an upper limit for two more. The average of the
two stars with measured Li is A(Li) = 0.63 from individual values with $\sigma$$_{Li}$ = 0.14. The two stars with only upper
limits have average limits of A(Li) = 0.45. These estimates are clearly below 1.25 and totally consistent with our clump
determinations obtained at lower resolution.

\acknowledgments

The authors are indebted to the referee for a thorough and invaluable reading of the original maunscript, identifying a number of potential points of confusion and a significant error in one of our equations. The paper has definitely been improved as a result. The authors gratefully acknowledge extensive use of the WEBDA\footnote{http:// webda.physics.muni.cz} database, maintained at the University of Brno by E. Paunzen, C. Stutz and J. Janik.  We also express appreciation to Jeffrey Cummings for sharing spectroscopic data with us. 
Support was provided to BJAT, DLB and BAT through NSF grant AST-1211621 and to CPD through AST-1211699.

\facility{WIYN:3.5m}
\software{IRAF}

\end{document}